\tolerance=10000
\magnification=\magstep1
\baselineskip=16pt
\vsize=7.5in
\hsize=5in
\hoffset=.25in
\ \vskip 1in
\centerline{\bf MASS AND WIDTH OF $\sigma$(750) SCALAR MESON FROM}
\centerline{\bf MEASUREMENTS OF $\pi N \to \pi^- \pi^+ N$ ON  
POLARIZED TARGETS}
\bigskip
\bigskip
\centerline{M. Svec\footnote*{Electronic address:\ \  
svec@hep.physics.mcgill.ca}}
\medskip
\centerline{Physics Department, Dawson College, Montreal, Quebec,  
Canada H3Z 1A4}
\smallskip
\centerline{and}
\smallskip
\centerline{McGill University, Montreal, Quebec, Canada H3A 2T8}
\vfill\eject
\noindent
{\bf Abstract.}
\medskip
The measurements of reactions $\pi^- p_\uparrow \to \pi^- \pi^+ n$  
at 17.2 GeV/c and $\pi^+ n_\uparrow \to \pi^+ \pi^- p$ at 5.98 and  
11.85 GeV/c made at CERN with polarized targets provide a  
model-independent and solution-independent evidence for a narrow  
scalar state $\sigma(750)$. The original $\chi^2$ minimization  
method and the recent Monte Carlo method for amplitude analysis of  
data at 17.2 GeV/c are in excellent agreement. Both methods find  
that the mass distribution of the measured amplitude $|\overline S  
|^2\Sigma$ with recoil transversity ``up'' resonates near 750 MeV  
while the amplitude $|S|^2\Sigma$ with recoil transversity ``down''  
is large and nonresonating. The amplitude $|S|^2\Sigma$ contributes  
as a strong background to $S$-wave intensity $I_S = (|S|^2 +  
|\overline S |^2)\Sigma$ and distorts the determinations of $\sigma$  
resonance parameters from $I_S$. To avoid this problem we perform a  
series of Breit-Wigner fits directly to the measured distribution  
$|\overline S |^2\Sigma$. The inclusion of various backgrounds  
causes the width of $\sigma(750)$ to become very narrow. Our best  
fit with $t$-averaged coherent background yields $m_\sigma = 753 \pm  
19$ MeV and $\Gamma_\sigma = 108 \pm 53$ MeV. These values are in  
excellent agreement with Ellis-Lanik theorem for the width of scalar  
gluonium. The gluonium interpretation of $\sigma(750)$ is also  
supported by the absence of $\sigma(750)$ in reactions $\gamma\gamma  
\to \pi\pi$. We also show how data on polarized target invalidate  
essential assumptions of past determinations of $\pi\pi$ phase  
shifts which explains the absence of $\sigma(750)$ in the  
conventional phase shift $\delta_0^0$. We examine the interference  
of $\sigma(750)$ with $f_0(980)$ and find it has only a very small  
effect on the determination of $\sigma(750)$ mass and width. The  
data on amplitude $|\overline S|^2\Sigma$ in the mass range of  
1120--1520 MeV show existence of a scalar resonance $f_0(1300)$ with  
a mass of 1280 $\pm$ 12 MeV and a width of 192 $\pm$ 26 MeV. We  
point out that the study of production processes on the level of  
spin amplitudes measured in experiments with polarized targets may  
reveal new hadron structures and open new physics beyond the  
standard QCD quark model.

\vfill\eject
\noindent
{\bf I. Introduction}
\medskip
Science is an ongoing interplay between our ideas and our  
experiences in the real Universe. We cannot expect to make a  
progress in our understanding unless we can show that what we  
thought we knew is in some sense incomplete or even wrong.$^1$  
Amplitude analyses of hadronic reactions provide a clear  
confirmation of this logic of scientific discovery.

In 1972, L.~van Rossum and his group at Saclay measured recoil  
nucleon polarization in $\pi^\pm p \to \pi^\pm p$ reactions$^2$ at  
CERN-PS. Their results closed the set of complete measurements of  
$\pi N \to \pi N$ reactions at 6 GeV/c and allowed them to make the  
first model independent determination of hadronic amplitudes  
directly from scattering data.$^3$ Failures of Regge models to  
correctly predict polarization and other spin observables were  
traced to the wrong structure of their amplitudes. Some Regge models  
were revised and constrained to reproduce the experimental $\pi N$  
amplitudes at 6 GeV/c.

In 1978 the Saclay group reported$^4$ the first amplitude analysis  
of $KN$ and $\overline K N$ charge exchange reactions, also at 6  
GeV/c. The structure of $A_2$-exchange amplitudes was found  
different from all revised Regge models. A more difficult amplitude  
analysis of $pp \to pp$ at 6 GeV/c was made possible by polarized  
proton beam at Argonne ZGS.$^5$ The results reported$^6$ in 1985 by  
A.~Yokosawa and his group at Argonne confirmed our lack of  
understanding of hadronic reactions at the level of amplitudes. So  
far, no new revisions of Regge models were even attempted. When the  
first measurement of a complete set of observables in elastic  
scattering of protons on ${}^{12}C$ carbon nucleus was reported$^7$  
in 1981, it invalidated the standard nonrelativistic analysis in  
favour of relativistic approach to nuclear physics.$^8$

The work by Saclay, Argonne and other groups firmly established the  
need for experimental knowledge of hadronic amplitudes. Efficient  
acquisition of this knowledge in two-body reactions has been  
hampered by the difficulty of measuring the recoil nucleon  
polarization. Dispersion relations were used in an effort to obtain  
hadronic amplitudes from incomplete data.$^9$ However, the situation  
is different in pion production reactions $\pi N \to \pi\pi N$ and  
$KN \to K\pi N$. In 1978, Lutz and Rybicki showed$^{10}$ that  
measurements of pion production in meson-nucleon scattering on  
transversely polarized target yield in a single experiment enough  
observables that almost complete and model independent amplitude  
analysis can be performed.

Amplitude analyses of pion production reaction such as $\pi N \to  
\pi\pi N$ or $NN\to\pi NN$ are important for two special reasons.  
First, these reactions provide information about unnatural exchange  
amplitudes which are not accessible in two-body reactions. Second,  
such amplitude analyses enable us to study resonance production on  
the level of spin-dependent amplitudes rather than spin-averaged  
cross-section $d^2\sigma/dmdt$.

The high statistics measurement of $\pi^- p \to \pi^- \pi^+ n$ at  
17.2 GeV/c at CERN-PS on unpolarized target$^{11}$ was later  
repeated with a transversely polarized proton target at the same  
energy$^{12{\rm --}16}$ Model independent amplitude analyses were  
performed for various intervals$^{12{\rm --}15}$ of dimeson mass of  
small momentum transfers $-t = 0.005$--0.2 (GeV/c)$^2$ and over a  
large interval$^{16}$ of momentum transfer $-t=0.2-1.0$ (GeV/c).$^2$

Additional information was provided by the first measurement of  
$\pi^+ n \to \pi^+ \pi^- p$ and $K^+ n \to K^+ \pi^- p$  
reactions$^{17, 18}$ on polarized deuteron target at 5.98 and 11.85  
GeV/c, also done at CERN-PS. The data allowed to study the  
$t$-evolution of mass dependence of moduli of amplitudes.$^{19}$  
Detailed amplitude analyses$^{20, 21}$ determined the mass  
dependence of amplitudes at larger momentum transfers $-t =  
0.2$--0.4 (GeV/c).$^2$

Amplitude analyses of $\pi^- p \to \pi^- \pi^+ n$ and $\pi^+ n \to  
\pi^+ \pi^- p$ reactions were recently repeated$^{22, 23}$ with  
special attention paid to error propagation and selection of  
physical solutions. This work was motivated by the emerging evidence  
from previous analyses for a new scalar resonance $\sigma(750)$.

All amplitude analyses$^{12-23}$ of pion production on polarized  
targets found a clear evidence for large and nontrivial unnatural  
$A_1$-exchange amplitudes in the dipion mass range from 400 to 1800  
MeV. This experimental finding is very important since previously  
the $A_1$ exchange amplitudes were assumed absent. In particular,  
all determinations of $\pi\pi$ phase shifts from unpolarized data on  
$\pi^- p \to \pi^- \pi^+ n$ are based on the assumption of  
vanishing $A_1$ exchange amplitudes.$^{24{\rm --}31}$ Without this  
assumption the determination of $\pi\pi$ phase shifts cannot even  
proceed.$^{23}$ The existence of large and nontrivial $A_1$ exchange  
amplitudes in $\pi N \to \pi^+ \pi^- N$ reactions casts a serious  
doubt about the reliability of the conventional $\pi\pi$ phase  
shifts. The assumption of absence of $A_1$ exchange amplitudes means  
that pion production in $\pi N \to \pi^+ \pi^- N$ does not depend  
on nucleon spin. What the measurements of $\pi N \to \pi^+ \pi^- N$  
on polarized targets found is that pion production depends strongly  
on nucleon spin. The dynamics of pion production is therefore not as  
simple as was assumed in the past determinations of $\pi\pi$ phase  
shifts.

The existence of $A_1$ exchange is also crucial for our  
understanding of spin structure of the nucleon. The measurements of  
the cross-section asymmetry using longitudinally polarized lepton  
beams on a longitudinally polarized nucleon targets determine  
nucleon spin dependent structure function $g_1 (x, Q^2)$. The  
measurements of $g_1$ on proton and neutron targets at CERN and SLAC  
has fascinating implications for the internal structure of the  
nucleon. These analyses of nucleon spin structure depend on the  
behaviour of $g_1$ for $x\to 0$ which is controlled by $A_1$  
exchange (see eq. (4.2.23) of Ref.~32).

Another important finding of measurements of $\pi^- p \to \pi^-  
\pi^+ n$ and $\pi^+ n \to \pi^+ \pi^- p$ reactions on polarized  
targets is the evidence for a narrow scalar state $I=0$  
$0^{++}(750)$.

The first published evidence for this state in $\pi^- p \to \pi^-  
\pi^+ n$ is presented in Fig.~2 of Ref.~14 from 1979. The  
CERN-Munich data on $S$-wave intensity $I_S$ show a clear bump  
around 750 or 800 MeV (depending on a solution). The CERN-Munich  
group refers to this bump as a resonance $\epsilon(800)$ but makes  
no Breit-Wigner fits. In 1979, Donohue and Leroyer published$^{33}$  
an analysis of the CERN-Munich data on polarized target and made the  
first claim that the data show existence of a narrow resonance  
which they called $\epsilon(750)$.

Our Saclay group measured $\pi^+ n \to \pi^+ \pi^- p$ on polarized  
deuteron target at 5.98 and 11.85 GeV/c at CERN--PS. The measured  
$S$-wave intensity $I_S$ showed a narrow resonant structure at 750  
MeV at both energies (see Fig.~10 of Ref.~20). To verify our results  
on $I_S$ in $\pi^+ n \to \pi^+ \pi^- p$ reaction, we analysed the  
CERN-Munich data at 17.2 GeV/c using our computer programs. All four  
solutions for $S$-wave intensity $I_S$ at all 3 energies showed a  
narrow resonant structure around 750 MeV. We reported these results  
in 1992 in Ref.~22 and made the claim for existence of a narrow  
scalar state $\sigma(750)$. However, later a numerical error was  
found in the final step of calculation of $S$-wave intensity $I_S$  
at 17.2 GeV/c. When corrected, only 2 solutions for $I_S$ at 17.2  
GeV/s show a clear narrow structure around 750 MeV while the other  
two solutions have a broader behaviour. Thus there appeared some  
difference between the results at low momentum transfer data at 17.2  
GeV/c and higher momentum transfer data at 5.98 and 11.85 GeV/c.

In Ref.~22 we presented analytical solutions of our amplitude  
analyses which included the unphysical solutions in many ($m,t$)  
bins. To deal with the unphysical solutions and to improve our error  
analysis we used a Monte Carlo method of amplitude analysis in  
Ref.~23. A clear signal for a narrow $\sigma(750)$ emerged from this  
improved analysis. This evidence for a narrow $\sigma(750)$ state  
was recently confirmed in a new measurement of $\pi^- p \to \pi^-  
\pi^+ n$ on polarized target at 1.78 GeV/c made at ITEP accelerator  
in Moscow and reported$^{34}$ in 1995.

To describe our new results, compare them to the CERN-Munich  
analyses, and to outline the program of this paper, we first need to  
introduce our notation and discuss the two accepted methods for  
amplitude analysis of data on polarized targets.

For masses below 1 GeV the dimeson system is produced predominantly  
in spin states $J=0$ ($S$-wave) and $J=1$ ($P$-wave). The  
experiments yield 15 spin-density-matrix (SDM) elements describing  
the dimeson angular distribution. These observables and the  
cross-section $d^2 \sigma/dmdt \equiv \Sigma$ can be expressed in  
terms of two $S$-wave and six $P$-wave nucleon transversity  
amplitudes.$^{10, 20}$ In our notation, the two $S$-wave amplitudes  
are $S$ and $\overline S$. The six $P$-wave amplitudes are $L,  
\overline L, U, \overline U$ and $N, \overline N$. The amplitudes  
$\overline A = \overline S, \overline L, \overline U, \overline N$  
and $A = S, L, U, N$ correspond to recoil nucleon transversity  
``up'' and ``down'' relative to the scattering plane. The amplitude  
analysis works with normalized amplitudes using a normalization
$$|S|^2 + |\overline S|^2 + |L|^2 + |\overline L|^2 + |U|^2 +  
|\overline U|^2 + |N|^2 + |\overline N|^2 = 1\eqno(1.1)$$
\noindent
The unnormalized amplitudes then are $|A|^2\Sigma$ and $|\overline  
A|^2\Sigma$. We also define partial-wave intensity
$$I_A = (|A|^2 + |\overline A |^2)\Sigma\eqno(1.2)$$
\noindent
where $A = S, L, U, N$. There are two solutions for the moduli  
$|A|^2$ and independent two solutions for the moduli $|\overline A  
|^2$. Hence, there are 4 independent solutions for the partial wave  
intensities $I_A$ which we label $I_A (i,j), i,j = 1,2$ with indices  
$i$ and $j$ referring to the two solutions for $|A|^2$ and  
$|\overline A|^2$, respectively.

Amplitude analysis expresses analytically$^{10, 20}$ the eight  
normalized moduli and the six cosines of relative phases of nucleon  
transversity amplitudes in terms of measured SDM elements. There are  
two similar solutions in each $(m, t)$ bin. However, in many  
$(m,t)$ bins the solutions are unphysical: either a cosine has  
magnitude larger than one or the two solutions for moduli are  
complex conjugate with a small imaginary part. Unphysical solutions  
also complicate the error analysis.

The occurrence of unphysical solutions is a common difficulty in  
all amplitude analyses. Two methods are used to find physical  
solutions and determine their errors. They are (a) $\chi^2$  
minimization method and (b) Monte Carlo method.

In the $\chi^2$ method one minimizes a function
$$\chi^2 = \sum\limits_{i=1}^M [{{{\rm Obs}_i {\rm (meas)} - {\rm  
Obs}_i {\rm (calc)}}\over{\Delta_i}}]^2\eqno(1.3)$$
\noindent
where ${\rm Obs}_i$ (meas) are the experimentally measured  
quantities, $\Delta_i$ are their experimental errors and ${\rm  
Obs}_i$ (calc) are corresponding expressions in terms of the  
amplitudes (moduli and cosines of relative phases). The analytical  
solutions for the moduli and cosines serve as initial values. This  
$\chi^2$ method was used in all CERN-Munich analyses$^{12-16}$ of  
$\pi^- p \to \pi^- \pi^+ p$ at 17.2 GeV/c. Since the two analytical  
solutions (initial values) are very close, the $\chi^2$ method leads  
to a unique solution in many $(m,t)$ bins. A particular exception  
is the mass range below 900 MeV. More recently the $\chi^2$ method  
was used in direct reconstruction of amplitudes of $pp$ elastic  
amplitudes from 0.8 to 2.7 GeV using polarized data obtained at  
SATURN II at Saclay.$^{35}$

The basic idea of Monte Carlo method is to vary randomly the input  
SDM elements within their experimental errors and perform amplitude  
analysis for each new set of the input SDM elements. The resulting  
moduli and cosines of relative angles are retained only when all of  
them have physical values in both analytical solutions. Unphysical  
solutions are rejected. The distributions of accepted moduli and  
cosines define the range of their physical values and their average  
value in each $(m,t)$ bin. The Monte Carlo amplitude analysis of  
Ref.~23 is based on 30,000 random variations of the input SDM  
elements. The Monte Carlo method was first used in 1977 in an  
amplitude analysis$^{36}$ of $pp$ elastic scattering at 6 GeV/c and  
later in an amplitude analysis$^{37}$ of reactions $\pi^- p \to K^+  
K^- n$ and $\pi^- p \to K_S^0 K_S^0 n$ at 63 GeV/c. In his review  
paper,$^{38}$ F.~James advocates the use of the Monte Carlo method  
as perhaps the only way to calculate the errors in the case of  
nonlinear functions which produce non-Gaussian distributions. The  
method has the added advantage that it can separate the physical and  
unphysical solutions and that it can retain the identity of the two  
analytical solutions.

The results for the two solutions for the unnormalized moduli of  
$S$-wave amplitudes $|\overline S |^2 \Sigma$ and $|S|^2\Sigma$  
obtained by Monte Carlo amplitude analysis of CERN-Munich data at  
17.2 GeV/c are shown in Fig.~1. We find that both solutions for the  
amplitude $|\overline S |^2\Sigma$ resonate around 750 MeV while  
both solutions for the amplitude $|S|^2\Sigma$ show non-resonant  
behaviour and increase with dipion mass $m$.

The results for $|\overline S |^2\Sigma$ and $|S|^2\Sigma$ obtained  
by $\chi^2$ minimization method using the same CERN-Munich data at  
17.2 GeV are shown in Fig.~2. We again find that both solutions for  
the amplitude $|\overline S |^2\Sigma$ resonate around 750 MeV and  
that both solutions for the amplitude $|S|^2\Sigma$ show  
non-resonant behaviour and increase with dipion mass $m$. The  
comparison of Fig.~1 and Fig.~2 shows that Monte Carlo method and  
$\chi^2$ minimization method are also in excellent numerical  
agreement. However, the amplitudes obtained by Monte Carlo method  
show a considerably smoother behaviour which, as we shall see later,  
gives much lower $\chi^2$ values in Breit-Wigner fits.

The unnormalized moduli $|\overline S |^2\Sigma$ and $|S|^2\Sigma$  
in Fig.~1 and 2 were calculated using $\Sigma = d^2 \sigma/dmdt$  
from Fig.~12 of Ref.~11.

At this point we note that the Fig.~2 is based on Fig.~10 of  
Ref.~13 and Fig.~VI--21 of Ref.~12 (to resolve error bars). The  
authors of these papers present only normalized moduli $|\overline  
S|$ and $|S|$ and consequently did not see the resonant behaviour of  
unnormalized amplitude $|\overline S|^2\Sigma$. The resonant  
behaviour of amplitude $|\overline S|^2\Sigma$ at 750 MeV went also  
unobserved in the subsequent analysis in Ref.~14 which was using  
polarized data in 40 MeV bins in the mass range from 600 to 1800  
MeV. It is possible to reconstruct the amplitudes $|\overline  
S|^2\Sigma$ and $|S|^2\Sigma$ from the information given in Ref.~14.  
As we shall see in Section V (Fig.~8) both solutions for  
$|\overline S|^2\Sigma$ resonate below 900 MeV while both solutions  
for $|S|^2\Sigma$ are nonresonating, in agreement with Figures 1 and  
2. It is interesting to note that the evidence for a narrow  
resonance $\sigma(750)$ was hidden in the very first analyses of  
CERN-Munich data (Ref.~12, 13 and 14) and was recognized by the  
present author some 16 years later in connection with the present  
work.

The aim of the present work is a more reliable determination of  
mass and width of $\sigma(750)$ resonance from model independent  
amplitude analyses of CERN-Munich data on $\pi^- p \to \pi^- \pi^+  
n$ on polarized target at 17.2 GeV/c. There are three important  
issues that we address in the process.

The first issue in the question which mass distribution should be  
used for Breit-Wigner fits to determine the resonance parameters of  
$\sigma(750)$ state. The previous CERN-Munich analyses$^{14-16}$  
fitted a Breit-Wigner formula to partial wave intensities, and we  
followed the same procedure in Ref.~23. However, the $S$-wave  
intensity at lower momentum transfers at 17.2 GeV/c shows a clear  
resonant structure only in solutions $I_S (1,1)$ and $I_S(2,1)$  
while the solutions $I_S(1,2)$ and $I_S(2,2)$ lack sufficient  
decreases of $I_S$ above 800 MeV to indicate a narrow resonance.  
This behaviour in $I_S = (|\overline S |^2 + |S|^2)\Sigma$ is caused  
by the large and nonresonating amplitude $|S|^2\Sigma$. The  
amplitude $|S|^2\Sigma$ thus behaves as a large and nonresonating  
background to the resonating amplitude $|\overline S |^2\Sigma$ and  
this distorts the determination of resonance parameters of  
$\sigma(750)$ from Breit-Wigner fits to $I_S$. To avoid this  
problem, it is necessary to perform Breit-Wigner fits directly to  
the resonant mass distributions $|\overline S |^2\Sigma$. Both  
solutions to $|\overline S |^2\Sigma$ resonate and the evidence for  
$\sigma(750)$ is thus entirely solution independent.

The second issue is which resonance shape formula is to be used in  
Breit-Wigner fits to $|\overline S |^2\Sigma$. The previous  
analyses$^{11,14{\rm --}16,23}$ used the Pi\v s\'ut-Roos shape  
formula which multiplies the standard Breit-Wigner formula (with a  
phase space) by an additional mass dependent factor $F = (2J + 1)  
(m/q)^2$.
\noindent
In their analysis of $\pi N \to \pi\pi N$ reaction  
amplitudes,$^{39}$  Pi\v s\'ut and Roos assumed absence of $A_1$  
exchange amplitudes and assumed that the mass dependence of pion  
production amplitudes is given by $\pi\pi$ scattering amplitudes.  
The partial wave expansion of $\pi\pi$ amplitudes then directly  
leads to the additional factor $F$. However, because of the  
existence of large and nontrivial $A_1$ exchange amplitudes in $\pi  
N \to \pi^+ \pi^- N$ reactions and because there is no proof that  
the mass dependence of $\pi N \to \pi\pi N$ production amplitudes is  
really described by the mass dependence of partial waves in  
$\pi\pi$ scattering, it is useful to perform Breit-Wigner fits to  
$|\overline S |^2\Sigma$ mass distribution using the standard$^{40}$  
phenomenological shape formula in which the  Pi\v s\'ut-Roos factor  
is absent (i.e. $F=1$) to see if there are differences in the  
determination of resonance parameters of $\sigma(750)$. The  
comparison of fits using both shape formulas finds only small  
differences.

The third issue is the question of background in the resonant mass  
distribution $|\overline S |^2\Sigma$. Nonresonating background  
comes e.g. from the isospin $I=2$ contribution to $|\overline S  
|^2\Sigma$. While it is difficult to exactly parametrize the unknown  
background, we estimated the background contribution using 3  
different approaches. In each case we find that inclusion of  
background leads to a significant reduction of the width of  
$\sigma(750)$ to somewhere around 100 MeV. Background is obviously  
important for the width determination of $\sigma(750)$ and thus to  
our understanding of the constituent structure of the $\sigma(750)$  
state. We also examine the interference of $\sigma(750)$ with  
$f_0(980)$ and find it has only a small effect on the mass and width  
of $\sigma(750)$.

The paper is organized as follows. In Section II we review the  
basic formalism. In Section III we derive the Pi\v s\'ut-Roos shape  
formula and describe the phenomenological shape formula for  
Breit-Wigner fits. In Section IV we present our fits to the measured  
resonating amplitude $|\overline S |^2\Sigma$ and describe our  
approaches to inclusion of coherent background. In Section V we  
study the interference of $\sigma(750)$ with $f_0(750)$ and also  
perform fits in the broad mass range 600--1520 MeV which show  
evidence for a scalar resonance $f_0(1300)$. In Section VI we  
present our fits to $S$-wave intensity $I_S$ in $\pi^- p \to \pi^-  
\pi^+ n$ at 17.2 GeV/c and in $\pi^+ n \to \pi^+ \pi^- p$ at 5.98  
and 11.85 GeV/c. In Section VII we review the assumptions of past  
determinations of $\pi\pi$ phase shifts and show how they are  
invalidated by the data on polarized targets. This explains the  
absence of narrow $\sigma(750)$ state in conventional phase shift  
$\delta_0^0$. In Section VIII we answer critical questions  
concerning the evidence for a narrow $\sigma(750)$. In Section IX we  
propose to identify the $\sigma(750)$ state with lowest mass  
gluonium $0^{++}(gg)$ and discuss theoretical and experimental  
support for this interpretation of $\sigma(750)$. In Section X we  
comment on the significance of studying hadron production on the  
level of spin amplituds and suggest that such studies may reveal new  
physics beyond the conventional QCD quark model of hadrons. The  
paper closes with a summary in Section XI.

\vfill\eject
\noindent
{\bf II. Basic Formalism}
\medskip
\noindent
A. \underbar{Phase space and amplitudes}.
\medskip
Various aspects of phase space, kinematics and amplitudes in pion  
production in $\pi N \to \pi\pi N$ reactions are described in  
several books.$^{41-44}$ In our discussion we will follow often the  
book by Pilkuhn.$^{41}$

Consider reaction $a+b\to 1+2+3$ such as $\pi^- p \to \pi^- \pi^+  
n$ with four-momentum conservation
$$P = p_a + p_b = p_1 + p_2 + p_3 = p_d + p_3\eqno(2.1)$$
\noindent
where $p_d = p_1 + p_2$ is the dipion momentum. The spin averaged  
cross-section is given by
$$d\sigma={1\over{{\rm Flux}(s)}}{1\over{(2s_a + 1)(2s_b + 1)}}  
\sum\limits_{\lambda_n,\lambda_p} | M_{\lambda_n,0\lambda_p}|^2 d  
{\rm Lips}_3\eqno(2.2)$$
\noindent
where the flux
$${\rm Flux}(s) = 4\sqrt{(p_a\cdot p_b)^2 - m_a^2 m_b^2} = 4M  
P_{\pi lab}\eqno(2.3)$$
\noindent
with $m_a = \mu$ mass of pion and $m_b=M$ mass of proton. The  
$\lambda_p$ and $\lambda_n$ are the proton and neutron helicities.  
The dipion state does not have a definite spin and helicity. The  
Lorentz invariant phase space is defined as
$$d {\rm Lips}_3 = (2\pi)^4 \delta^4 (P-p_1 - p_2 - p_3)  
\prod\limits_{i=1}^3 {{d^3 p_i}\over{(2\pi)^3(2E_i)}}\eqno(2.4)$$
\noindent
We will work with the usual kinematic variables of c.m. energy  
squared $s$, momentum transfer $t$, dipion mass $m$ and angles  
$\theta,\phi$ describing the angular distribution of $\pi^-$ in the  
$\pi^- \pi^+$ rest frame. Hence
$$\eqalign{s & = (p_a + p_b)^2 = (p_1 + p_2 + p_3)^2\cr
t & = (p_a - p_d)^2 = (p_a - p_1 - p_2)^2\cr
m^2 & = p_d^2 = (p_1 + p_2)^2\cr}\eqno(2.5)$$
\noindent
Following the procedure described on pp.~18--19 of Ref.~40 and  
using $dm^2 = 2m dm$, we can write
$$d{\rm Lips}_3 (P, p_1, p_2, p_3) = q(m^2) G(s) dmdtd\Omega$$
\noindent
where $q$ is the pion momentum in the c.m.s. of the dipion system
$$q(m^2) = \sqrt{0.25 m^2 - \mu^2} = {m\over 2} \sqrt{1-  
({{2\mu}\over m})^2}\eqno(2.6a)$$
\noindent
and the energy dependent part of phase space
$$G(s) = {1\over{(2\pi)^4}} {1\over{8\sqrt{\lambda (s,\mu^2,  
M^2)}}}\eqno(2.6b)$$
\noindent
where
$$\lambda(s,\mu^2, M^2) = [s - (\mu + M)^2] [s-(\mu - M)^2]$$
\noindent
Hence
$${{d\sigma}\over{dmdtd\Omega}} = {{K(s)}\over{4\pi}} q  
\sum\limits_{\lambda_p, \lambda_n} |M_{\lambda_n, 0 \lambda_p} (s,t,  
m,\theta,\phi)|^2\eqno(2.7)$$
\noindent
where
$$K(s) = {{2\pi G(s)}\over{{\rm Flux}(s)}}$$
\noindent
To obtain dipion states of definite spin $J$ and helicity  
$\lambda$, we expand
$$M_{\lambda_n, 0\lambda_p} = \sum\limits_{J=0}^\infty  
\sum\limits_{\lambda = -J}^{+J} \sqrt{(2J + 1)} M^J_{\lambda  
\lambda_n, 0\lambda_p} (s,t,m)d^J_{\lambda 0}  
(\theta)e^{i\lambda\phi}\eqno(2.8)$$
\noindent
We now integrate $|M_{\lambda_n, 0\lambda_p}|^2$ over $d\Omega$.  
Using orthogonality relations for the $d$-functions and spherical  
harmonics, we obtain for the reaction cross-section
$${{d^2\sigma}\over{dmdt}} = q(m^2) K(s) \sum\limits_{J=0}^\infty  
\sum\limits_{\lambda, \lambda_n, \lambda_p} |M^J_{\lambda \lambda_n,  
0 \lambda_p} (s,t,m)|^2\eqno(2.9)$$
\noindent
We now define
$$H^J_{\lambda \lambda_n, 0\lambda_p} = \sqrt{q(m^2)} \sqrt{K(s)}  
M^J_{\lambda \lambda_n, 0\lambda_p}\eqno(2.10)$$
\noindent
We will also consider only the $J=0$ ($S$-wave) and $J=1$  
($P$-wave) contributions for dipion masses $m$ below 1000 MeV. With  
a notation
$$\Sigma = d^2\sigma/dmdt\eqno(2.11)$$
\noindent
we now define normalized helicity amplitudes with definite  
$t$-channel naturality
$$\eqalign{H^0_{0+, 0+} & = S_0 \sqrt\Sigma,\quad H^0_{0+,0-} = S_1  
\sqrt\Sigma\cr
H^1_{0+,0+} & = L_0 \sqrt\Sigma ,\quad  H^1_{0+,0-} = L_1 \sqrt\Sigma\cr
H^1_{\pm 1 +,0+} & = {1\over{\sqrt 2}} (N_0 \pm U_0)\sqrt\Sigma\cr
H^1_{\pm 1+, 0-} & = {1\over{\sqrt 2}}(N_1 \pm U_1)  
\sqrt\Sigma\cr}\eqno(2.12)$$
\noindent
In (2.12) $n = |\lambda_p - \lambda_n | = 0,1$ is the nucleon  
helicity flip. At large $s$, the unnatural helicity nonflip  
amplitudes $S_0, L_0, U_0$ and the unnatural helicity flip  
amplitudes $S_1, L_1, U_1$ exchange $\pi$ and $A_1$ quantum numbers  
in the $t$-channel, respectively. Both natural exchange amplitudes  
$N_0$ and $N_1$ exchange $A_2$ at large $s$.

The amplitude analysis of data on polarized targets is performed  
using normalized recoil nucleon transversity amplitudes defined as
\bigskip
\tabskip=1em plus2em minus.5em
\halign to \hsize{#\hfil & #\hfil & \hfil #\cr
$\displaystyle S = {1\over{\sqrt 2}} (S_0 + i S_1)$ &  
$\displaystyle \overline S = {1\over{\sqrt 2}} (S_0 - i S_1)$\cr
$\displaystyle  L = {1\over{\sqrt 2}} (L_0 + i L_1)$ &  
$\displaystyle  \overline L = {1\over{\sqrt 2}} (L_0 - i L_1)$\cr
$\displaystyle U = {1\over{\sqrt 2}} (U_0 + i U_1)$ &  
$\displaystyle  \overline U = {1\over{\sqrt 2}} (U_0 - i U_1)$  
&\hfill (2.13)\cr
$\displaystyle N = {1\over{\sqrt 2}} (N_0 - i N_1)$ &  
$\displaystyle \overline N = {1\over{\sqrt 2}} (N_0 + i N_1)$\cr}
\bigskip
The amplitudes $S, L, U, N$ and $\overline S, \overline L,  
\overline U, \overline N$ correspond to recoil nucleon transversity  
``down'' and ``up'', respectively.$^{18,20}$ The ``up'' direction is  
the direction of normal to the scattering plane defined according  
to Basel convention by ${\vec p}_\pi \times {\vec p}_{\pi\pi}$ where  
${\vec p}_\pi$ and ${\vec p}_{\pi\pi}$ are the incident pion and  
dimeson momenta in the target nucleon rest frame.

The normalized amplitudes satisfy conditions
$$\sum\limits^1_{n=0} |S_n|^2 + |L_n|^2 + |U_n|^2 + |N_n|^2 = 1$$
$$|S|^2 + |\overline S |^2 + |L|^2 + |\overline L |^2 + |U|^2 +  
|\overline U |^2 + |N|^2 + |\overline N |^2 = 1\eqno(2.14)$$
\noindent
We now define spin-averaged partial wave intensities for amplitudes  
$A = S, L, U, N$
$$I_A = (|A_0|^2 + |A_1|^2) \Sigma = (|A|^2 + |\overline A |^2)  
\Sigma\eqno(2.15)$$
\noindent
Obviously
$$\Sigma = {{d^2\sigma}\over{dmdt}} = I_S + I_L + I_U + I_N\eqno(2.16)$$
>From the point of view of Breit-Wigner fits to various mass  
distributions in $\pi^- p \to \pi^- \pi^+ n$ $(\Sigma, I_A,  
|\overline S |^2 \Sigma$ etc.) the important point is that the part  
of the phase space which depends on the dipion mass $m$ is simply  
the c.m. pion momentum $q$ given by (2.6a).
\bigskip
\noindent
{\bf B. Spin observables and amplitude analysis.}
\medskip
For invariant masses below 1000 MeV, the dipion system in reactions  
$\pi N \to \pi^+ \pi^- N$ is produced predominantly in spin states  
$J=0$ ($S$-wave) and $J=1$ ($P$-wave). The experiments on  
transversely polarized targets then yield 15 spin-density-matrix  
(SDM) elements describing the dipion angular distribution. The  
measured SDM elements are$^{17, 18}$
$$\rho_{ss} + \rho_{00} + 2\rho_{11}, \rho_{00} - \rho_{11},  
\rho_{1-1},\eqno(2.17a)$$
$${\rm Re} \rho_{10}, {\rm Re} \rho_{1s}, {\rm Re} \rho_{0s},$$
$$\rho_{ss}^y + \rho_{00}^y + 2\rho_{11}^y, \rho_{00}^y -  
\rho_{11}^y, \rho_{1-1}^y,\eqno(2.17b)$$
$${\rm Re} \rho_{10}^y, {\rm Re} \rho_{1s}^y, {\rm Re} \rho_{0s}^y,$$
$${\rm Im}\rho_{1-1}^x, {\rm Im} \rho_{10}^x, {\rm Im}  
\rho_{1s}^x.\eqno(2.17c)$$
\noindent
The SDM elements (2.17a) are also measured in experiments on  
unpolarized targets. The observables (2.17b) and (2.17c) are  
determined by the transverse component of target polarization  
perpendicular and parallel to the scattering plane $\pi N \to (\pi^+  
\pi^-)N$, respectively. The SDM elements (2.17) depend on $s,t$,  
and $m$. There are two linear relations among the matrix elements in  
(2.17):
$$\eqalignno{\rho_{ss} + \rho_{00} + 2\rho_{11} & = 1,\cr
\rho_{ss}^y + \rho_{00}^y + 2\rho_{11}^y & = A,&(2.18)\cr}$$
\noindent
where $A$ is the polarized target asymmetry.

The data analysis is carried out in the $t$-channel helicity frame  
for the $\pi^+ \pi^-$ dimeson system. The helicities of the initial  
and final nucleons are always defined in the $s$-channel helicity  
frame. Using the notation of (2.17) and (2.18), the first group of  
equations represents the sum of SDM elements (2.17a) and (2.17b):
$$\eqalignno{a_1 & = {1\over 2} [1+A] = |S|^2 + |L|^2 + |U|^2 +  
|\overline N|^2,\cr
a_2 & = [(\rho_{00} - \rho_{11}) + (\rho_{00}^y - \rho_{11}^y)] =  
2|L|^2 - |U|^2 - |\overline N|^2,\cr
a_3 & = [\rho_{1-1} + \rho_{1-1}^\nu ] = |\overline N|^2 -  
|U|^2,&(2.19a)\cr
a_4 & = {1\over{\sqrt 2}} [{\rm Re}\rho_{10} + {\rm Re} \rho_{10}^y  
] = |U| |L| \cos (\gamma_{LU}),\cr
a_5 & = {1\over{\sqrt 2}} [{\rm Re} \rho_{1s} + {\rm Re}  
\rho_{1s}^y ] = |U| |S| \cos (\gamma_{SU}),\cr
a_6 & = {1\over 2} [{\rm Re} \rho_{0s} + {\rm Re} \rho_{0s}^y ] =  
|L| |S| \cos (\gamma_{SL}).&(2.19b)\cr}$$
\noindent
Similar equations relate the difference of SDM elements to  
amplitudes of opposite transversity. The second group of observables  
is defined as
$$\eqalignno{{\overline a}_1 & = {1\over 2} [1-A] = |\overline S|^2  
+ |\overline L|^2 + |\overline U|^2 + |N|^2,\cr
{\overline a}_2 & = [(\rho_{00} - \rho_{11}) - (\rho_{00}^y -  
\rho_{11}^y)] = 2 |\overline L|^2 - |\overline U|^2 - |N|^2,\cr
{\overline a}_3 & = [\rho_{1-1} - \rho_{1-1}^y] = |N|^2 -  
|\overline U|^2,&(2.20a)\cr
{\overline a}_4 & = {1\over{\sqrt 2}} [{\rm Re}\rho_{10} - {\rm  
Re}\rho_{10}^y ] = |\overline U| |\overline L| \cos  
({\overline\gamma}_{LU}),\cr
{\overline a}_5 & = {1\over{\sqrt 2}} [{\rm Re}\rho_{1s} - {\rm  
Re}\rho_{1s}^y ] = |\overline U| |\overline S| \cos  
({\overline\gamma}_{SU}),\cr
{\overline a}_6 & = {1\over 2} [{\rm Re}\rho_{0s} - {\rm  
Re}\rho_{0s}^y ] = |\overline L| |\overline S| \cos  
({\overline\gamma}_{SL}),&(2.20b)\cr}$$
\noindent
In Eqs.~(2.19b) and (2.20b) we have introduced explicitly the  
cosines of relative phases between the nucleon transversity  
amplitudes.

The SDM elements (2.17c) form the third group of observables,$^{10,  
20}$ which is not used in the present amplitude analysis.

The first group (2.19) involves four moduli $|S|^2$, $|L|^2$,  
$|U|^2$ and $|\overline N|^2$ and three cosines of relative phases  
$\cos (\gamma_{SL})$, $\cos(\gamma_{SU})$, and $\cos(\gamma_{LU})$.  
The second group (2.20) involves the same amplitudes, but with  
opposite nucleon transversity. Analytical solution for these  
amplitudes in terms of observables was derived in Ref.~10 and 20.  
For the first group one obtains a cubic equation for $|L|^2 \equiv  
x$:
$$ax^3 + bx^2 + cx + d = 0,\eqno(2.21)$$
\noindent
with coefficients $a,b,c,d$ expressed in terms of observables $a_i,  
i = 1,2,\ldots , 6$. The remaining moduli and the cosines are given  
by the expressions
$$\eqalignno{|S|^2 & = (a_1 + a_2) - 3|L|^2,\cr
|U|^2 & = |L|^2 - {1\over 2} (a_2 + a_3),\cr
|\overline N|^2 & = |L|^2 - {1\over 2} (a_2 - a_3),\cr
\cos(\gamma_{LU}) & = {{a_4}\over{|L| |U|}},\cr
\cos(\gamma_{SU}) & = {{a_5}\over{|S| |U|}},\qquad  
\cos(\gamma_{SL}) = {{a_6}\over{|S| |L|}}.&(2.22)\cr}$$
\noindent
The solution for the second group (2.20) is similar.

The analytical solutions of the cubic equation (2.21) are given in  
Table I of Ref.~20. One solution of (2.21) is always negative and it  
is rejected. The other two solutions are generally positive and  
close. However, in a number of $(m,t)$ bins we get unphysical values  
for some cosines and in some cases also negative moduli of  
amplitudes. In some $(m,t)$ bins the mean values of input SDM  
elements yield complex solutions for $|L|^2$ or $|\overline L|^2$ or  
both (with positive real parts). To filter out the unwanted  
unphysical solutions and to determine the errors on the amplitudes  
and their average values, one can use either the $\chi^2$  
minimization method or the Monte Carlo method. The results of  
amplitude analyses of $\pi^- p \to \pi^- \pi^+ n$ at 17.2 GeV/c for  
dipion masses in the range 600--900 MeV are given in Ref.~13 for the  
$\chi^2$ minimization method and in Ref.~23 for the Monte Carlo  
method. The two methods are in excellent agreement. In Ref.~23 we  
also present the Monte Carlo amplitude analysis of the reaction  
$\pi^+ n \to \pi^+ \pi^- p$ at 5.98 and 11.85 GeV/c using the Saclay  
data$^{17}$ at larger momentum transfers $-t = 0.2 - 0.4$  
(GeV/c)$^2$ and dipion masses in the range 360--1040 MeV.
\bigskip
\noindent
{\bf III. Resonance shape formulas.}
\medskip
\noindent
{\bf A. Pi\v s\'ut-Roos shape formula.}
\medskip
Before we review the Pi\v s\'ut-Roos derivation of their resonance  
shape formula, we first recall some properties of partial waves in  
elastic scattering of scalar particles. The $T$-matrix amplitude of  
isospin I has partial wave expansion
$$T^I (s,\cos\theta) = 8\pi \sum\limits_{L=0}^\infty (2L + 1) T_L^I  
(s) P_L (\cos\theta)\eqno(3.1)$$
\noindent
The unitarity in elastic scattering then requires (see Ref.~40,  
pp.~38--40) that $T^I_L$ has a form
$$T_L^I(s) = {{\sqrt s}\over q} \sin\delta_L^I e^{i\delta_L^I} =  
{{\sqrt s}\over q} {1\over{\cot\! {\rm g} \delta_L^I -  
i}}\eqno(3.2)$$
\noindent
where $q$ is the pion c.m.s. momentum and $\delta_L^I$ is the  
corresponding phase shift. Notice that the factor $\sqrt s/q$ is  
induced by the unitarity alone. At a resonance $m_R$, the  
relativistic Breit-Wigner formula for $T_L^I$ then reads $$T_L^I =  
{{\sqrt s}\over q} {{- m_R \Gamma(s)}\over{(s-m_R^2) + im_R  
\Gamma(s)}}\eqno(3.3)$$
\noindent
where $\Gamma(s)$ is an energy dependent width.

Let us now return to pion production process $\pi^- p \to \pi^-  
\pi^+ n$ and to amplitudes $M^J_{\lambda\lambda_n, 0\lambda_p}  
(s,t,m)$ defined in (2.8). In their analysis,$^{39}$ Pi\v s\'ut and  
Roos assumed that the following amplitudes vanish for all $J$:
$$M^J_{0+,0+} = 0\eqno(3.4a)$$
$$M^J_{\pm1+,0+} = M^J_{\pm 1 +, 0-} = 0\eqno(3.4b)$$
\noindent
The conditions (3.4) mean that all $A_1$-exchange amplitudes vanish  
and that the natural $A_2$-exchange amplitudes also vanish. Only  
pion exchange amplitude $M^J_{0+,0-}$ contribute and they have a  
general form
$$M^J_{0+,0-} (s,t,m) =  Q (s,t) \sqrt{2J + 1} T^J (m) \sqrt{f(m)}  
+ M^J_B (s,t,m)\eqno(3.5)$$
\noindent
where $T^J (m)$ are the $\pi\pi \to \pi\pi$ partial wave amplitudes  
with isospin decomposition
$$T^J = T^J_{I=1}\ \hbox{for}\ J \ \hbox{odd}\eqno(3.6)$$
$$T^J = {2\over 3} T^J_{I=0} + {1\over 3} T^J_{I=2}\ \hbox{for}\ J\  
\hbox{even}$$
\noindent
in reaction $\pi^+ \pi^- \to \pi^+ \pi^-$. In (3.5) the function  
$f(m)$ is a phenomenological function that is supposed to account  
for absorption, and initial-state and final-state interactions. In  
practice one puts $f(m) = 1$. The function $Q(s,t)$ factorizes the  
$s$- and $t$-dependence. The term $M^J_B (s,t,m)$ is a background.

Taking into account the factor $qK(s)$ in (2.9) and the equation  
(3.3), Pi\v s\'ut and Roos arrive at a resonant parametrization of  
reaction cross-section
$${{d^2\sigma}\over{dmdt}} = q (2J + 1) ({m\over q})^2 {{m^2_R  
\Gamma^2}\over{(m^2 - m^2_R)^2 + m^2_R \Gamma^2}} f(m) {\cal  
N}(s,t)\eqno(3.7)$$
$$+\ \hbox{background terms}$$
\noindent
Averaging over $t$ over an interval $<t_1, t_2>$ gives a shape  
formula for the mass distribution
$$I(s,m) = q (2J + 1) ({m\over q})^2 {{m^2_R \Gamma^2}\over{(m^2 -  
m^2_R)^2 + m^2_R \Gamma^2}} f(m) N(s)\eqno(3.8)$$
$$+\ \hbox{background terms}$$
\noindent
where
$$N(s) = {1\over{t_2 - t_1}} \int\limits_{t_1}^{t_2} {\cal N} (s,t)  
dt = {{K(s)}\over{t_2 - t_1}} \int\limits_{t_1}^{t_2} | Q (s,t) |^2  
dt\eqno(3.9)$$
\noindent
Setting $f(m) =1$ and ignoring the background we get the Pi\v  
s\'ut-Roos resonance shape formula$^{39}$ for $t$-averaged mass  
distribution
$$I(m) = N q F(m) |BW|^2\eqno(3.10)$$
\noindent
where $N$ is the normalization constant, $q$ is the phase space  
factor, $F(m)$ is the Pi\v s\'ut-Roos shape factor
$$F(m) = (2J + 1) ({m\over q})^2 =  
{{4(2J+1)}\over{1-({{2\mu}\over{m}})^2}}\eqno(3.11)$$

\noindent
and $BW$ is the Breit-Wigner amplitude
$$BW = {{m_R \Gamma}\over{m^2_R - m^2 - i m_R \Gamma}}\eqno(3.12)$$
\noindent
The Pi\v s\'ut-Roos resonance shape formula (3.10) has been  
extensively used to fit partial wave intensities in previous  
amplitude analyses of $\pi N \to \pi^+ \pi^- N$ on polarized targets  
(Ref.~14, 15, 16 and 23).
\bigskip
\noindent
{\bf B. Phenomenological resonance shape formula.}
\medskip
In general, the experimental distribution $I(m)$ in a certain mass  
region is fitted to a functional form$^{40}$
$$I(m) = \alpha_R I_R (m, m_R, \Gamma) + \alpha_B I_B (m)\eqno(3.13)$$
\noindent
where $\alpha_R$ and $\alpha_B$ give the fractions of resonant  
contribution and incoherent background. Normally $I_R$ is taken as a  
square of the Breit-Wigner amplitude multiplied by a phase space  
factor. A coherent term may be added to the Breit-Wigner amplitude,  
typically a constant term. In general, the background $I_B(m)$ is a  
polynomial.

In the case of $\pi^- p \to \pi^- \pi^+ n$ reaction, the relevant  
phase space factor is just the pion momentum $q$ in the $\pi^+  
\pi^-$ c.m. system and one can write for mass distributions in this  
reaction a phenomenological resonance shape formula
$$I(m) = N q(m) \{ |BW|^2 + B\}\eqno(3.14)$$
\noindent
where $N$ is overall normalization factor and $B$ is the background  
term. We can take $B=0$ or $B=$constant. When $B=0$, the  
phenomenological shape formula (3.14) is obtained from Pi\v  
s\'ut-Roos resonance shape formula (3.10) by setting their shape  
factor $F\equiv 1$. We see from (3.11) that Pi\v s\'ut-Roos formula  
(3.10) converges to phenomenological formula (3.14) for large $m$  
when background $B=0$.
\bigskip
\noindent
{\bf IV. The mass and width of $\sigma(750)$ from fits to $S$-wave  
amplitude $|\overline S|^2\Sigma$.}
\medskip
As seen in Fig.~1 and 2, the Monte Carlo method and the $\chi^2$  
method yield very similar results for the $S$-wave amplitudes  
$|\overline S|^2\Sigma$ and $|S|^2\Sigma$ in $\pi^- p \to \pi^-  
\pi^+ n$ at 17.2 GeV/c and for $-t = 0.005 - 0.20$ (GeV/c)$^2$. Both  
methods show that the amplitude $|\overline S|^2\Sigma$ resonates  
in both solutions while the amplitude $|S|^2\Sigma$ is  
non-resonating in both solutions. The Monte Carlo results appear to  
be smoother than the $\chi^2$ results. The Monte Carlo method found  
no physical solution at mass bin 890 MeV. The solution found by  
$\chi^2$ method at this mass is far off from the general trend of  
data in solution 1 for $|\overline S|^2\Sigma$. For these reasons  
the mass bin 880--900 MeV was excluded from the fits to $|\overline  
S |^2\Sigma$.

To determine the best values of the mass and width of $\sigma(750)$  
state from the mass distribution of the resonating amplitude  
$|\overline S |^2\Sigma$ we used 4 types of fitting approaches and  
used a $\chi^2$ criterion to determine the best fits. In the first  
approach we used a single Breit-Wigner fit. In the second approach  
we added an incoherent constant background to the single  
Breit-Wigner. In the third and fourth approaches we used two  
different versions of constant coherent background. In each approach  
we used both Pi\v s\'ut-Roos and phenomenological resonance shape  
formula and found they give very similar results. The inclusion of  
background leads to the narrowing of the width of $\sigma(750)$. The  
best $\chi^2$ solution is obtained by the fourth approach leading  
to a conclusion that $\sigma(750)$ is a narrow state with a width  
about 100 MeV. The fitting was done using the CERN optimization  
program FUMILI.$^{45}$
\bigskip
\noindent
{\bf A. Single Breit-Wigner fit.}
\medskip
In this approach the mass distribution $|\overline S |^2\Sigma$ is  
fitted to a formula
$$|\overline S |^2\Sigma = q FN_S |BW|^2\eqno(4.1)$$
\noindent
where $q$ is the phase space factor (2.6a). The factor $F$ is equal  
either
$$F = (2J + 1) ({m\over q})^2\eqno(4.2a)$$
\noindent
for Pi\v s\'ut-Roos shape formula or
$$F=1.\eqno(4.2b)$$
\noindent
for the phenomenological shape formula. $BW$ is the Breit-Wigner  
amplitude
$$BW = {{m_R\Gamma}\over{m^2_R - m^2 - i m_R\Gamma}}\eqno(4.3)$$
\noindent
where $m_R$ is the resonant mass. The mass dependent width $\Gamma  
(m)$ depends on spin $J$ and has a general form
$$\Gamma = \Gamma_R ({q\over{q_R}})^{2J+1} {{D_J (q_R r)}\over{D_J  
(qr)}}\eqno(4.4)$$
\noindent
In (4.4) $q_R = q (m = m_R)$ and $D_J$ are the centrifugal barrier  
functions of Blatt and Weishopf:$^{46}$
$$\eqalignno{D_0 (qr) & = 1.0\cr
D_1  (qr) & = 1.0 + (q r)^2&(4.5)\cr}$$
\noindent
where $r$ is the interaction radius.

The results of the fit are shown in Fig.~3 and 4 for the Pi\v  
s\'ut-Roos and phenomenological shape formulas, respectively. The  
corresponding curves for both shape formulas are nearly identical.  
The numerical results are presented in Table 1. The fits to  
$|\overline S |^2\Sigma$ obtained by $\chi^2$ method have  
significantly higher values of $\chi^2$/d.o.f. However both methods  
give a $\sigma$ mass in the range 730--750 MeV and a width in the  
range 230--250 MeV. Only the solution 1 of the $\chi^2$ method gives  
a lower width around 190 MeV.

An important feature of the fits to $|\overline S |^2\Sigma$ with  
single Breit-Wigner formula noticeable in Fig.~3 and 4 is that all  
fits lie below the maximum values of the mass distributions for each  
solution and the method of analysis. This inability of the single  
Breit-Wigner formula to reproduce the resonant shape of the  
amplitude $|\overline S |^2\Sigma$ suggests that background  
contributions are important and their effect on the mass and width  
of $\sigma$ state should be investigated, at least approximatively.

\bigskip
\noindent
{\bf B. Breit-Wigner fit with incoherent background.}
\medskip
In this case we fit the mass distribution for $|\overline S  
|^2\Sigma$ to a formula
$$|\overline S |^2\Sigma = q FN_S \{ |BW|^2 + B\}\eqno(4.6)$$
\noindent
where $B$ is the incoherent background added to the Breit-Wigner  
formula (4.1). In general, $B$ is a polynomial in $m$. However,  
since we have only 14 data points in the resonant mass range of  
600--880 MeV, we will take $B=$constant.

The results of the fit are shown in Fig.~5 for the phenomenological  
shape formula ($F=1$). The results with Pi\v s\'ut-Roos shape  
formula are very similar. The numerical results are given in Table  
2. We notice a dramatic improvement of the fit to the solution 2 for  
both methods which yields a better $\chi^2$/dof and a narrow width  
of about 100 MeV. There is also some improvement of the fit to the  
solution 1 in particular for the $\chi^2$ method solution. This  
improvement is again associated with a lower $\chi^2$/dof and a  
narrower width of 202 MeV and 147 MeV for the Monte Carlo and  
$\chi^2$ methods, respectively. The mass of the $\sigma$ state  
remains in the range of 730--745 MeV.

While the fits to solutions 2 are now much improved, the fits to  
solutions 1 are still not satisfactory with the fitted curves still  
below the maximum values of these mass distributions. To make  
further progress we turn to coherent background contributions.

\bigskip
\noindent
{\bf C. Breit-Wigner fit with coherent background.}
\medskip
The nonresonant behaviour of the amplitude $|S|^2\Sigma$ (recoil  
nucleon transversity down) strongly suggest the presence of a  
coherent nonresonating background. A part of coherent background  
also comes from the contribution of isospin $I=2$ amplitudes (see  
eq.~(3.6)) which we neglected in the single Breit-Wigner fit. To  
understand the origins of the coherent background and to discuss its  
form for fits to $|\overline S |^2\Sigma$ it is useful to express  
the unnormalized moduli of $S$-wave transversity amplitudes in terms  
of unnormalized helicity amplitudes. Using (2.13) we write
$$\eqalignno{|S|^2\Sigma & = {1\over 2} |S_0 + iS |^2\Sigma = q F  
|F_0 + iF_1 |^2\cr
|\overline S |^2\Sigma & = {1\over 2} |S_0 - iS_1 |^2\Sigma = qF |  
F_0 - iF_1 |^2&(4.7)\cr}$$
\noindent
where $F_0$ and $F_1$ are unnormalized $S$-wave helicity  
amplitudes. The terms $qF$ have the same meaning as in (4.1) and  
anticipate the use of (4.7) for Breit-Wigner fits to mass  
distribution of $|\overline S |^2\Sigma$. Near the resonance with  
mass $m_R$ we assume the following form of the helicity amplitudes
$$F_n (s,t,m) = R_n (s,t,m) BW(m) + B_n (s,t,m)\eqno(4.8)$$
\noindent
where $n=0,1$ is the nucleon helicity flip, $BW$ is the  
Breit-Wigner amplitude (4.3), $R_n(s,t,m)$ is the pole term and  
$B_n(s,t,m)$ is the nonresonating background which includes the  
contribution from the nonresonating isospin $I=2$ amplitudes. The  
energy variable $s$ is fixed and will be omitted in the following.  
Since the experimental mass distributions are averaged over broad  
$t$-bins, we will eventually average also over the momentum transfer  
variable $t$. With the notation $\epsilon = \pm 1$, we can then  
write (4.7) in a compact form as follows
$$F_ 0 + i\epsilon F_1 = R_\epsilon (t,m) BW(m) + B_\epsilon  
(t,m)\eqno(4.9)$$
\noindent
where $R_\epsilon = R_0 + i\epsilon R$, and $B_\epsilon = B_0 +  
i\epsilon B$. It is useful to factor out the phase of $R_\epsilon$  
and define
$$\eqalignno{R_\epsilon & = |R_\epsilon | e^{i\phi_\epsilon}\cr
C_\epsilon & = B_\epsilon e^{-i\phi_\epsilon}&(4.10)\cr}$$
\noindent
Then (4.9) takes the form
$$F_0 + i\epsilon F_1 = \{ |R_\epsilon | BW + C_\epsilon \}  
e^{i\phi_\epsilon}\eqno(4.11)$$
\noindent
and the moduli squared of (4.7) read
$$|F_0 + i\epsilon F_1|^2 = |R_\epsilon|^2 |BW|^2 + ({\rm Re}  
C_\epsilon)^2 + ({\rm Im}  C_\epsilon)^2 +$$
$$ + 2 |R_\epsilon | \{ {\rm Re} C_\epsilon {\rm Re} BW +    {\rm  
Im} C_\epsilon {\rm Im} BW \}\eqno(4.12)$$
\noindent
We now recall that
$$\eqalignno{ {\rm Re} BW & = ({{m_R - m^2}\over{m_R \Gamma}})  
|BW|^2 \equiv w |BW|^2\cr
{\rm Im} BW & = |BW|^2&(4.13)\cr}$$
\noindent
Hence
$$|F_0 + i\epsilon F_1 |^2 = \{ | R_\epsilon |^2 + 2 |R_\epsilon|  
{\rm Re} C_\epsilon w + 2 |R_\epsilon | {\rm Im} C_\epsilon\}  
|BW|^2$$
$$+ ({\rm Re} C_\epsilon)^2 + ({\rm Im} C_\epsilon)^2\eqno(4.14)$$
\noindent
Since the amplitude $|S|^2\Sigma (\epsilon = +1)$ does not show a  
clear resonant behaviour (Fig.~1 and 2), we can conclude from (4.14)  
that the sum of terms in the parentheses must be small or zero.  
This most likely means that $|R_+|$ is small or zero implying that  
the pole terms in helicity amplitudes are related approximately as  
$R_0 \approx -iR_1$.

For the resonating amplitude $|\overline S |^2\Sigma (\epsilon=-1)$  
the second and third terms in the parentheses in the equation  
(4.14) represent the effect of coherent background. In general the  
functions $|R_-|$ and $C_-$ will depend on both $t$ and $m$. Since  
these functions are not known and since we have only 14 data points  
in the resonance mass region 600--880 MeV, we will work in the  
approximation of constant background. At this point there are two  
possibilities.

\noindent
(1) We assume that $|R_-|$ and $C_-$ are constants independent of  
$t$ and $m$. In this case no averaging over $t$ is necessary and we  
can write (4.14) in the form
$$|\overline S |^2\Sigma = q F N_S \{ [ 1 + 2 w B_1 + 2 B_2] |BW|^2  
+ B_1^2 + B_2^2 \}\eqno(4.15)$$
\noindent
where
$$N_S = |R_-|^2,\quad B_1 = {{{\rm Re} C_-}\over{|R_-|}},\quad B_2  
= {{{\rm Im} C_-}\over{|R_-|}}\eqno(4.16)$$
\noindent
This possibility is equivalent to assuming that the constant parts  
of $|R_-|$ and $C_-$ dominate in the resonant mass range 600--880  
MeV. We also notice that in this case the incoherent part $B_1^2 +  
B_2^2$ is correlated with the coherent contribution on $(2w B_1 +  
2B_2) |BW|^2$ in the formula (4.16) through the common parameters  
$B_1$ and $B_2$.
\medskip
\noindent
(2) In the second possibility, we assume that $|R_-|$ and $C_-$ are  
both dependent on $t$ and $m$. In this case we must average (4.14)  
over $t$ over the experimentally measured interval $<t_1 , t_2 >$.  
The averaging of (4.14) over $t$ yields
$$|\overline S |^2\Sigma = q F \{ [r + 2   wa + 2b] |BW|^2 +  
c\}\eqno(4.17)$$
\noindent
where
$$\eqalignno{r & = <|R_-|^2>,\  a = < |R_-| {\rm Re} C_->\cr
b & = <|R_-| {\rm Im} C_- > ,\  c = <({\rm Re}C_-)^2 + ({\rm Im}  
C_-)^2>&(4.18)\cr}$$
\noindent
In (4.18) the symbol $<\ >$ represents averaging over $t$ over  
interval $<t_1, t_2>$. In general, the functions $r, a, b, c$ will  
depend on the mass $m$. Since we do not know these functions, we  
will assume constant values. But then there is no distinction  
between $r$ and $2b$ which can be combined into one parameter $N_S =  
r + 2b$ as they are two constants in a sum. Then (4.17) has the  
form
$$|\overline S|^2\Sigma = q FN_S \{[1 + 2w B_1] |BW|^2 + B\}\eqno(4.19)$$
\noindent
where $B_1 = b/N_S$ and $B = c/N_S$ are the coherent and incoherent  
contributions to the resonance shape formula. This approximation is  
equivalent to assumption that the functions $|R_-|$ and $C_-$  
depend mostly on $t$ and only weakly on $m$. Notice that in this  
case the incoherent contribution $B$ is not correlated with the  
coherent contribution as the parameters $B$ and $B_1$ are  
independent.

We will refer to the first possibility (1) as Breit-Wigner fit with  
constant coherent background and to the second possibility (2) as  
the Breit-Wigner fit with $t$-averaged constant coherent background.

The results of the Breit-Wigner fit with constant coherent  
background are shown in Fig.~6 and Table 3. The results of the  
Breit-Wigner fit with the $t$-averaged constant coherent background  
are given in Fig.~7 and Table 4. Both Figures and Tables refer to  
the phenomenological shape formula with $F=1$. The results with Pi\v  
s\'ut-Roos resonance shape formula ($F$ given by (4.2a)) are very  
similar for the masses and widths although there are some  
differences in the fitted values of the constants $B_1, B_2$ or  
$B_1$ and $B$.

An inspection of Figures 6 and 7 show much improved fits to the  
data on mass distribution of $|\overline S |^2\Sigma$. The overall  
best fit (as judged by the lowest values of $\chi^2$/dof) is  
provided by the Breit-Wigner fit with the $t$-averaged constant  
coherent background. However the improvements in $\chi^2$/dof appear  
only in solution 1 of Monte Carlo method and solution 2 of the  
$\chi^2$ method. Again, the Monte Carlo method achieves beter values  
of $\chi^2$/dof compared to the $\chi^2$ method of amplitude  
analysis.

The improvements in the fits brought about by the inclusion of  
coherent background have important consequences for the fitted  
values of the mass and width of $\sigma(750)$ state. From Tables 3  
and 4 we find that the mass of $\sigma$ in solution 1 is about 30  
MeV higher than the $\sigma$ mass found in solution 2. The Monte  
Carlo method gives the best value of $\sigma$ mass 774 MeV in  
solution 1 and 744 MeV in solution 2 (Table 4). The $\chi^2$ method  
gives the best value of $\sigma$ mass 761 MeV in solution 1 and 733  
MeV in solution 2 (Table 4). The data on polarized target cannot  
distinguish these two solutions. Since the two masses are close, we  
can work with a solution average. The solution average for $\sigma$  
mass is 759 $\pm$ 22 MeV for Monte Carlo method and 747 $\pm$ 16 MeV  
for the $\chi^2$ method. The average over the two methods gives  
$\sigma$ mass 753 $\pm$ 19 MeV.

The most significant effect of the inclusion of coherent background  
is the reduction of the value of the width of $\sigma$. The Monte  
Carlo method gives for the best value of $\sigma$ width similar  
values of 101 MeV and 103 MeV in solution 1 and 2, respectively  
(Table 4). The $\chi^2$ method gives for the best value of $\sigma$  
width 134 MeV in solution 1 and 93 MeV in solution 2 (Table 4). The  
data on polarized target cannot distinguish these two solutions, but  
the high values of $\chi^2$/dof for $\chi^2$ method tend to favour  
the values for $\sigma$ width from the Monte Carlo method which has  
low values of $\chi^2$/dof. The solution average for the $\sigma$  
width is 102 $\pm$ 61 MeV for Monte Carlo method. The solution  
average for the $\sigma$ width is 113 $\pm$ 44 MeV for the $\chi^2$  
method. Since the error on the $\sigma$ width is larger for the  
Monte Carlo method, the two results are essentially compatible. The  
average over the two methods gives $\sigma$ width 108 $\pm$ 53 MeV.

In conclusion, we propose to adopt the solution and method averages  
from the best fit values of Table 4 as the standard values of mass  
and width of the $\sigma$ state. The obtained values are
$$m_\sigma = 753 \pm 19\ \hbox{MeV}\quad ,\quad \Gamma_\sigma = 108  
\pm 53\ \hbox{MeV}\eqno(4.20)$$

\bigskip
\noindent
{\bf V. The interference with $f_0 (980)$ in fits to amplitude  
$|\overline S |^2\Sigma$}
\medskip
T\"ornqvist suggested$^{47}$ that the interference of $\sigma(750)$  
with $f_0(980)$ resonance could influence the determination of  
resonance parameters of $\sigma(750)$. In the old phase shift  
analyses (obtained using the invalid assumption of absence of  
$A_1$-exchange), the resonance $f_0(980)$ plays an important role of  
smoothly interpolating the ``Down'' solution for $\delta_0^0$ below  
900 MeV with the results for $\delta_0^0$ above 1000 MeV.

We will now investigate the effect of interference of $\sigma(750)$  
with $f_0(980)$ on the determination of resonance parameters of  
$\sigma(750)$. We will find that the effect is very small. This is  
consistent with the fact that $f_0(980)$ is a very narrow resonance  
and it is positioned sufficiently far away from the narrow and  
strong resonance $\sigma(750)$.

The experimental data in the $f_0(980)$ mass region are given in  
the CERN-Munich analysis$^{14}$ of $\pi^- p \to \pi^- \pi^+ n$ on  
polarized target at 17.2 GeV/c for dipion masses 600--1800 MeV. From  
Fig.~2 and Fig.~6 of Ref.~14 it is possible to reconstruct the  
amplitudes $|\overline S |^2\Sigma$ and $|S|^2\Sigma$. The two  
solutions are shown in Fig.~8. The amplitude $|\overline S  
|^2\Sigma$ resonates at 750 MeV in solution 1 and at 800 MeV in  
solution 2. It shows a high value at 960 MeV and a pronounced dip at  
1000 MeV, indicating an interference of $f_0(980)$ with background  
in this mass region around 1000 MeV. The structures are less  
dramatic in $|S|^2\Sigma$ which does not show $\sigma(750)$ but a  
dip at 1000 MeV is still observable.

To proceed, we extend our parametrization (4.8) of $|\overline S  
|^2\Sigma$ to include $f_0(980)$ resonance. Recall from (4.7) that  
$|\overline S |^2\Sigma = q F | F_0 -i F_1|$. Now we write for the  
helicity amplitudes $F_0$ and $F_1$
$$F_n = F_n^{(\sigma)} (s,t,m) BW_\sigma (m) + R_n^{(f)} (s,t,m)  
BW_f (m) + B_n (s,t,m)\eqno(5.1)$$
\noindent
where index $\sigma$ refers to $\sigma(750)$ and $f$ refers to  
$f_0(980)$.

\noindent
Then
$$F_0 - i F_1 = R_\sigma (s,t,m) BW_\sigma + R_f (s,t,m) BW_f +  
B(s,t,m)\eqno(5.2)$$

\noindent
Assuming that the coefficients $R_\sigma, R_f$ and the background  
$B$ are independent of $t$ and $m$, we get an extension of the  
parametrization (4.15)
$$|\overline S |^2\Sigma = q F N_S \{ [ 1 + 2w_\sigma B_1 + 2 B_2]  
| BW_\sigma|^2+\eqno(5.3)$$
$$+ B_1^2 + B_2^2 + [C_1^2 + C_2^2 ] |BW_f|^2 +$$
$$+2 [ (w_\sigma |BW_\sigma|^2 + B_1) (w_f C_1 - C_2)+$$
$$+ (|BW_\sigma|^2 + B_2) (C_1 + w_f C_2)] |BW_f|^2\}$$
\noindent
where
$$w_R = {{m_R^2 - m^2}\over{m_R \Gamma}}\qquad ,\qquad \Gamma =  
\Gamma_R ({q\over{q_R}})\qquad , \qquad R = \sigma,f\eqno(5.4)$$
If we assume that $R_\sigma, R_f$ and $B$ depend on $t$ and perform  
$t$-averaging, the extension of parametrization (4.19) then reads
$$|\overline S |^2 \Sigma = q FN_S\{[ 1 + 2w_\sigma B_1]  
|BW_\sigma|^2 + B+$$
$$+ 2 [w_\sigma B_1 + w_\sigma (w_f C_1 - C_2) |BW_f|^2 +$$
$$+ (C_1 + w_f C_2) |BW_f|^2 ] |BW_\sigma |^2+$$
$$+ (D_1 + w_f D_2) |BW_f|^2\}\eqno(5.5)$$
In the above parametrizations (5.3) and (5.5) the coefficients  
$N_S$, $B_1$, $B_2$, $(B)$, $C_1$, $C_2$, $D_1$, $D_2$ are real  
constants. The data between 900 and 1120 MeV exist only in 40 MeV  
mass bins. Thus there is not enough data to fit the resonance  
parameters of $f_0(980)$. Instead we fix the mass of $f_0(980)$ at  
980 MeV and its width at 48 MeV in the Breit-Wigner amplitude  
$BW_f$. Also, in our fits we took for $|\overline S |^2\Sigma$ below  
880 MeV the results from our Monte Carlo analysis (in 20 MeV bins)  
and between 900 and 1120 MeV we took the results of CERN-Munich  
analysis (in 40 MeV bins) from Fig.~8.

The two parametrizations (5.3) and (5.5) yield virtually identical  
fits from 600 to 1120 MeV and the same values for mass and width of  
$\sigma(750)$. The fit for parametrization (5.3) is shown in Fig.~9  
and the numerical values of the parameters are given in Table 5.  
There is a small improvement of $\chi^2$/dof in Solution 1 which  
shows a better fit with the $f_0(980)$ interference. Comparison with  
the corresponding Table 3 shows a small increase in the mass of  
$\sigma$ in both solutions. There is a decrease of the $\sigma$  
width in solution 1 from 114 MeV to 95 MeV and an increase in  
$\sigma$ width in Solution 2 from 104 MeV to 135 MeV. The solution  
averages are
$$m_\sigma = 768\pm 22\ \hbox{MeV}\qquad , \qquad \Gamma_\sigma =  
115\pm 38\ \hbox{MeV}\eqno(5.6)$$
\noindent
The effect of $f_0(980)$ interference is thus a small increase of  
average mass and width of $\sigma$ as compared to values in (4.20).  
It is not possible to claim$^{47}$ that the low mass and the narrow  
width of $\sigma(750)$ are artifacts due to the neglect of  
interference of $\sigma(750)$ with $f_0(980)$ in our fits.

Both fits reproduce well the $\sigma$ resonance peaks below 880 MeV  
in both solutions and the interference patterns between 920 and  
1120 MeV. Particularly noteworthy in Fig.~9 is the dramatic drop in  
$|\overline S |^2\Sigma$ between 960 and 1000 MeV due to destructive  
interference of $f_0(980)$ with the background. The good fit in  
this region suggests that the assumption of constant coherent  
background and resonance couplings is a good approximation.

We have also attempted to fit the whole mass region of 600--1520  
MeV using a three resonance parametrization with a constant  
background and resonance couplings. The fit was not successful as  
the $f_0(1300)$ resonance was not well reproduced. This indicates  
that the background above 1120 MeV is different and the assumption  
of constant background for such a large mass range does not work.  
Next we fitted the $f_0(1300)$ resonance in the mass range of 1120  
to 1520 MeV to a single Breit-Wigner with incoherent background. The  
results are shown in Fig.~9 for the two solutions (differing in  
values of $|\overline S |^2\Sigma$ at 1480 MeV). The Solution 1  
above 1120 MeV connects smoothly with both solutions below 1120 MeV  
while the Solution 2 shows a small discontinuity at 1120 MeV.  
Surprisingly, the incoherent background in both solutions is  
consistent with zero. This again indicates that above 1100--1200 MeV  
the background (if any) is different from the low mass region below  
1100 MeV. The numerical results of the fit to $f_0(1300)$ in the  
mass region 1120--1520 MeV are given in Table 6. We note the  
similarity of mass and width of resonances $f_0(1300)$ and  
$f_2(1270)$.

\bigskip
\noindent
{\baselineskip=12pt
{\bf VI. The mass and width of $\sigma(750)$ state from the fits to  
$S$-wave }\break

\noindent
{\qquad \bf intensity $I_S$.}\par}
\medskip
Previous amplitude analyses$^{13-15, 23}$ of $\pi^-  
p\to\pi^-\pi^+n$ and $\pi^+                 n \to \pi^+ \pi^- p$  
data on polarized targets fitted only certain partial wave  
intensities using Pi\v s\'ut-Roos resonance shape formula without  
any background. It is of interest to perform Breit-Wigner fits to  
the $S$-wave intensity $I_S$ and compare the results with the  
results of fits to resonating amplitude $|\overline S |^2\Sigma$ in  
$\pi^- p \to \pi^- \pi^+ n$ at 17.2 GeV/c. Because of lower  
statistics, data for $\pi^+ n \to \pi^+ \pi^- p$ at 5.98 and 11.85  
GeV/c allow fits only to $S$-wave intensity $I_S$. This is thus our  
primary aim in fitting $S$-wave intensity: to extract information  
about the mass and width of $\sigma$ in $\pi^+       n\to \pi^+\pi^-  
p$ reaction measured at larger momentum transfers $-t = 0.2 - 0.4$  
(GeV/c)$^2$.

Let us recall that the $S$-wave intensity $I_S$ is defined as
$$I_S (s,t,m) = (|S|^2 + |\overline S |^2)\Sigma  = (|S_0|^2 +  
|S_1|^2)\Sigma\eqno(6.1)$$
\noindent
Since there are two independent solutions for the amplitudes  
$|S|^2$ and $|\overline S |^2$, there are 4 solutions for the  
$S$-wave intensity. We label these 4 solutions as $I_S(1,1),  
I_S(1,2), I_S(2,1)$ and $I_S(2,2)$ where
$$I_S (i,j) = (|S(i)|^2 + |\overline S (j) |^2)\Sigma\quad ,\quad  
i,j = 1,2\eqno(6.2)$$
\noindent
The results for the 4 solutions of $I_S$ obtained by the Monte  
Carlo amplitude analysis are shown in Fig.~10. The results for  
$I_S(1,1)$ and $I_S(2,2)$ obtained by the $\chi^2$ minimization  
method are shown in Fig.~11. Again, there is a remarkable agreement  
between the results of these two different methods of analysis. The  
solutions $I_S(1,1)$ and $I_S(2,1)$ are clearly resonating but the  
solutions $I_S(1,2)$ and $I_S(2,2)$ do not show a clear resonant  
behaviour. This is due to the large nonresonating contribution from  
the amplitude $|S|^2\Sigma$ (see Fig.~1 and 2). The amplitude  
$|S|^2\Sigma$ represents a nontrivial nonresonating background in  
all four solutions and is thus expected to distort the results of  
Breit-Wigner fits to $I_S$.

We first performed fits to $I_S$ using a single Breit-Wigner  
formula without any background
$$I_S = q F N_S |BW|^2\eqno(6.3)$$
\noindent
In all fits to $S$-wave intensities we used the Pi\v s\'ut-Roos  
shape factor $F=(2J+1) (m/q)^2$. The results are shown as solid  
lines in Fig.~10 and 11 and in Tables 7 and 8 for Monte Carlo and  
$\chi^2$ methods, respectively.  We notice in Fig.~10 and 11 that  
the single Breit-Wigner fit is well below the maximum values of the  
mass distribution $I_S$. In Monte Carlo analysis the mass of  
$\sigma$ is around 766 MeV in all four solutions. The width is  
around 260 MeV for the first 3 solutions and is larger at 303 MeV  
for the solutions $I_S(2,2)$. In $\chi^2$ method the $\sigma$ mass  
and width in solution $I_S(1,1)$ is in agreement with Monte Carlo  
results, but the width of $I_S(2,2)$ is larger at 408 MeV and also  
mass is higher at 786 MeV.

Next we performed Breit-Wigner fit with a constant incoherent  
background using a formula
$$I_S = q F N_S \{ |BW|^2 + B\}\eqno(6.4)$$
\noindent
where $B$ is the constant background term. The results are shown as  
dashed lines in Fig.~10 and 11 and in Tables 7 and 8 for the Monte  
Carlo and $\chi^2$ methods, respectively. While the masses of  
$\sigma$ remain the same, there is a general reduction of the width  
of $\sigma$ associated with improved fits to the data and lower  
values of $\chi^2$/dof. In Monte Carlo method the width of $\sigma$  
is reduced to 210 MeV in the first 3 solutions to $I_S$. However the  
most dramatic and unexpected change occurs in the solution  
$I_S(2,2)$ in both methods. There is a considerable improvement in  
the fit to the data and the width is drastically reduced to 188 MeV  
in both methods indicating the existence of a narrow $\sigma$ state  
even in the broad looking mass distribution.

The best determination of $\sigma$ width from the fits to $S$-wave  
intensity $I_S$ is still double of the best  value obtained in fits  
directly to the amplitude $|\overline S |^2\Sigma$ (Table 4). This  
discrepancy shows that the determination of resonance parameters  
from the spin-averaged intensities is not fully reliable when there  
is a presence of large nonresonating nontrivial background as is the  
case of the amplitude $|S|^2\Sigma$. The characteristic feature of  
this situation is that the $S$-wave intensity does not show a clear  
reonant structure in all four solutions.

This situation does not occur in the data on $S$-wave intensity in  
$\pi^+ n \to \pi^+ \pi^- p$ at larger momentum transfers $-t =  
0.2-0.4$ (GeV/c)$^2$. The results from Monte Carlo amplitude  
analysis are shown in Fig.~12 and 13 at 5.98 and 11.85 GeV/c,  
respectively. We notice that all four solutions at both energies  
show clear resonant structures. This suggests that the determination  
of resonance parameters from $S$-wave intensities at these momentum  
transfers should be more reliable. However, this advantage is  
somewhat offset by the lower statistics of the data and large  
errors.

We have again performed fits using single Breit-Wigner formula  
(6.3) and the Breit-Wigner fit with constant incoherent background  
using formula (6.4). The results are shown in Fig.~12 and 13 and in  
Tables 9 and 10 for incident momenta of 5.98 and 11.85 GeV/c,  
respectively. The fit with constant background (dashed lines) is a  
clear improvement over a single Breit-Wigner fit (solid lines). The  
improvement of the fit with the constant background is again  
associated with lower values of $\chi^2$/dof and with reduction of  
the width of $\sigma$ in all solutions at both energies. However,  
there are differences in values for the mass and the width of  
$\sigma$ between the solutions as well as between energies. At 5.98  
GeV/c, the mass ranges from 706 to 745 MeV and the width ranges from  
145 to 262 MeV. At 11.85 GeV/c, the mass is higher and ranges from  
756 to 782 MeV while the width is lower ranging from 117 to 202 MeV.  
The differences are probably due to lower statistics.

The solution averages for the mass and width of $\sigma$ from fits  
to $I_S$ are as follows:

\noindent
At 5.98 GeV/c
$$m_\sigma  = 730\ {\rm MeV} \pm 27\ {\rm MeV}\quad ,\quad  
\Gamma_\sigma = 195 \pm 81\ {\rm MeV}\eqno(6.5)$$
\noindent
At 11.85 GeV/c
$$m_\sigma = 768 \pm 17\ {\rm MeV}\quad ,\quad \Gamma_\sigma = 166  
\pm 54\ {\rm MeV}\eqno(6.6)$$
\noindent
At 17.2 GeV/c
$$m_\sigma = 767 \pm 9\ {\rm MeV}\quad ,\quad \Gamma_\sigma = 204  
\pm 75\ {\rm MeV}\eqno(6.7)$$
\noindent
The best values for the mass and width of $\sigma$ obtained from  
fits to the $S$-wave intensities at the three energies are in  
general agreement. The small differences are likely due to the fact  
that the approximation of constant incoherent background may work  
differently at various energies and momentum transfers. The  
differences in mass of $\sigma$ from the fits to $|\overline S  
|^2\Sigma$ and to $I_S$ are small. The difference in the value of  
the width from the best fits to $|\overline S |^2\Sigma$ with  
coherent background and the fits to $I_S$ are somewhat large but the  
results are still consistent. At 17.2 GeV/c they are due to large  
nonresonating contributions from the amplitude $|S|^2\Sigma$. The  
differences also reflect the need for inclusion of coherent  
background and its better description than a constant. This in turn  
would require more data of high statistics in the resonance region  
600--900 MeV.

\bigskip
\noindent
{\bf VII. Remarks on determinations of $\pi\pi$ phase shifts.}
\medskip
The amplitude analyses of measurements of $\pi N_\uparrow \to \pi^+  
\pi^- N$ on polarized targets provide a model-independent and  
solution-independent evidence for a narrow scalar state $I=0\  
0^{++}(750)$. The question arises how to understand the absence of  
such a state in the conventional $S$-wave phase shift $\delta_0^0$  
in $\pi\pi$ scattering.$^{11,24-31}$

Of course, there are no actual measurements of pion-pion scattering  
and there is no partial-wave analysis of $\pi\pi\to\pi\pi$  
reactions in the usual sense. The $\pi\pi$ phase shifts are  
determined indirectly from measurements of $\pi^- p \to \pi^- \pi^+  
n$ on unpolarized targets using several strong enabling assumptions.  
One of these crucial assumptions -- the absence of $A_1$ exchange  
amplitudes -- leads to predictions for polarized spin density matrix  
(SDM) elements and for the measured amplitudes, and it is thus  
directly testable in the measurements on polarized targets. As we  
shall see below, the assumption of absence of $A_1$-exchange  
amplitudes is totally invalidated by the data on polarized targets.  
The polarization measurements also cast some doubt on the  
fundamental assumption of factorization of mass $m$ and momentum  
transfer $t$ in the crucial pion exchange amplitudes. We must use  
the results of measurements on polarized targets to judge the  
validity of $\pi\pi$ phase shifts, and not vice versa. We are thus  
led to the conclusion that the indirect and model-dependent  
determinations of $\pi\pi$ phase shifts cannot be correct. This  
explains the absence of $I=0\ 0^{++}(750)$ resonance in the  
$\delta_0^0$ phase shift from these analyses.

We will now review the basic assumptions common to all  
determinations of $\pi\pi$ phase shifts.$^{11,24-31}$

A priori, there is no connection between the partial wave  
amplitudes in $\pi\pi\to\pi\pi$ scattering and the production  
amplitudes in $\pi N \to \pi^+ \pi^- N$ reactions. We recall that in  
$\pi N \to \pi^+ \pi^- N$ there are two $S$-wave production  
amplitudes $S(s,m,t)$ and $\overline S (s,m,t)$ (or $S_0 (s,m,t)$  
and $S_1 (s,m,t)$) while in $\pi\pi\to\pi\pi$ there is one $S$-wave  
amplitude (or phase shift $\delta_0^0$) dependent only on the energy  
$E$. Also, in $\pi N \to \pi^+ \pi^- N$ there are six $P$-wave  
production amplitudes $L, \overline L, U, \overline U, N, \overline  
N$ (or $L_n, U_n, N_n, n = 0,1$) which depend on variables $s,m,t$  
while in $\pi\pi\to\pi\pi$ there is again one $P$-wave amplitude (or  
phase shift $\delta_1^1$) dependent only on the energy $E$. To make  
the connection between the production amplitudes in $\pi N \to  
\pi^+ \pi^- N$ and the partial-wave amplitudes in $\pi\pi\to\pi\pi$  
the following assumptions of factorization and identification are  
postulated in all determinations of $\pi\pi$ phase shifts from  
unpolarized data on $\pi N \to \pi^+ \pi^- N$.

The starting point are the dimeson helicity $\lambda=0$ pion  
exchange amplitudes $S_1$ and $L_1$ in the $t$-channel. It is  
assumed that the $t$ and $m$ dependence in these amplitudes  
factorizes:
$$\eqalignno{S_1 (s_,m,t) & = N {{\sqrt{-t}}\over{t-\mu^2}} F_0 (t)  
{m\over{\sqrt q}} f_0 (m)\cr
L_1(s,m,t) & = N {{\sqrt{-t}}\over{t-\mu^2}} F_1 (t) {m\over{\sqrt  
q}} f_1 (m) & (7.1)\cr}$$
\noindent
where $t$ is the momentum transfer at the nucleon vertex, $m$ and  
$q$ are the dipion mass and the $\pi^-$ momentum in the $\pi^-\pi^+$  
c.m. frame. The form factors $f_J(t)$ describe the $t$-dependence  
and the functions $f_J(m)$, $J=0,1$, describe the mass dependence.  
$N$ is a normalization constant. Furthermore, the functions $f_J(m)$  
are assumed to be the partial-wave amplitudes in $\pi^-\pi^+ \to  
\pi^-\pi^+$ reaction at c.m. energy $m$:
$$\eqalignno{ f_0 & = {2\over 3} f_0^{I=0} + {1\over 3} f_0^{J=2}\cr
f_1 & = f_1^{I=1} & (7.2)\cr}$$
\noindent
The partial wave amplitudes $f_J^I$ with definite isospin $I$ are  
defined so that in the $\pi\pi$ elastic region
$$f_J^I = \sin \delta_J^I e^{i\delta_J^I}\eqno(7.3)$$
\noindent
The phase shifts $\delta_J^I$ are determined from the amplitudes  
$S_1$ and $L_1$ which are calculated from the data on $\pi^- p \to  
\pi^- \pi^+ n$ on unpolarized target. However the calculation of  
amplitudes $S_1$ and $L_1$ from the $\pi^- p \to \pi^- \pi^+ n$ data  
cannot be done without additional assumptions. There is simply more  
amplitudes than data. To proceed further all determinations of  
$\pi\pi$ phase shifts must assume that all $A_1$-exchange amplitudes  
vanish:
$$S_0 = L_0 = U_0 \equiv 0\eqno(7.4)$$
\noindent
With the assumptions (7.4), two solutions for the $S$-wave phase  
shift $\delta_0^0$ are found:$^{27,28}$ ``Down'' solution which is  
non-resonating and ``Up'' solution which resonates at the mass  
around 770 MeV with a width about 150 MeV. The resonating solution  
was rejected because it disagreed with the $\pi^0 \pi^0$ mass  
spectrum from a low-statistics experiment$^{48}$ on $\pi^- p \to  
\pi^0 \pi^0 n$ at 8 GeV/c.

There is no theoretical proof of factorization (7.1) and  
identification (7.2) of functions $f_J$ with $\pi\pi$ partial-wave  
amplitudes. It is not obvious that the $\pi\pi$ phase shifts  
calculated from $\pi^- p \to \pi^- \pi^+ n$ data using the  
assumptions (7.1)--(7.3) would coincide with $\pi\pi$ phase shifts  
determined directly from real pion-pion scattering. Only such  
comparison could test the assumption (7.2).

The factorization (7.1) implies that the mass spectrum of  
amplitudes $|S_1|^2$ and $|L_1|^2$ is independent of $t$. This  
consequence of factorization can be tested in measurements of $\pi N  
\to \pi^+ \pi^- N$ on polarized targets. In Fig.~14 we show  
$t$-evolution of mass dependence of lower and upper bounds$^{19}$ on  
normalized moduli $|L|^2$, $|\overline L |^2$, $|U|^2$ and  
$|\overline U |^2$. The data at $t=-0.068$ (GeV/c)$^2$ are from  
$\pi^- p \to \pi^-\pi^+ n$ at 17.2 GeV/c, the rest is from $\pi^+ n  
\to \pi^+ \pi^- p$ at 5.98 GeV/c. The Fig.~14 shows a clear and  
pronounced dependence of mass spectra of amplitudes $|L|^2$ and  
$|\overline L|^2$ on momentum transfer $t$. In particular, there is  
a clear change of mass spectrum below $-t = 0.25$ (GeV/c)$^2$, a  
region of $t$ relevant to determinations of $\pi\pi$ phase shifts.  
While this change could be entirely due to $A_1$ exchange amplitude  
$L_0$, this cannot be guaranteed. The factorization assumption (7.1)  
thus cannot be taken for granted and further tests of this  
assumption are required in future high statistics measurements of  
$\pi N \to \pi^+ \pi^- N$ on polarized targets.

The assumption (7.4) of absence of $A_1$-exchange amplitudes has  
several consequences that can be directly tested in measurements on  
polarized targets. >From (2.13) we see that absence of $A_1$  
exchange amplitudes implies
$$|A| = |\overline A | \qquad \hbox{for}\qquad A = S, L, U\eqno(7.5)$$
\noindent
The equality of moduli of amplitudes with the recoil nucleon  
transversity ``down'' and ``up'' is not observed experimentally. We  
can see in Fig.~1 and 2 that the $S$-wave amplitudes $|S|$ and  
$|\overline S |$ are clearly unequal at 17.2 GeV/c and $-t =  
0.005-0.20$ (GeV/c)$^2$. In Fig.~14 we see that the $P$-wave  
amplitudes $|L|$ and $|\overline L |$ are different in every $t$-bin  
from 0.005 to 0.60 (GeV/c)$^2$, and that the difference is largest  
at small $t$, the region of most importance to determination of  
$\pi\pi$ phase shifts.

The $A_1$-exchange is large and nontrivial also above 900 MeV and  
in higher partial waves $D$ and $F$. This finding of CERN-Munich  
analysis$^{14}$ of $\pi^- p \to \pi^- \pi^+ n$ data on polarized  
target in the mass range 600--1800 MeV is shown in Fig.~15. The  
figure shows the ratios of moduli of amplitudes with recoil nucleon  
transversity ``down'' and ``up'' for $S$-, $P$-, $D$- and $F$-wave  
amplitudes with dimeson helicity $\lambda=0$ which are directly  
relevant for determination of the corresponding phase shifts. The  
deviations from 1 indicate the strength of $A_1$-exchange. We can  
see in Fig.~15 that $A_1$-exchange is important in all waves up to  
1800 MeV at small $-t = 0.005-0.20$ (GeV/c)$^2$. The determinations  
of $\pi\pi$ phase shifts above 900 MeV also assumed the absence of  
$A_1$-exchange amplitudes. We must conclude that the determinations  
of $\pi\pi$ phase shifts from $S$-wave to $F$-wave in the mass  
region from 600 to 1800 MeV are not reliable. Theoretical  
calculations and analyses based on these phase shifts are therefore  
not reliable as well.

Below 1000 MeV, where the $S$- and $P$-wave dominate, the  
assumptions (7.4) lead to predictions for polarized SDM elements  
that can be directly compared with the data. The predictions of  
(7.4) are$^{23}$
$$\rho^y_{ss} + \rho^y_{00} + 2\rho^y_{11} = - 2 (\rho^y_{00} -  
\rho^y_{11}) = + 2 \rho^y_{1-1}\eqno(7.6)$$
$${\rm Re} \rho^y_{10} = {\rm Re} \rho^y_{1s} = {\rm Re}  
\rho^y_{0s} \equiv 0\eqno(7.7)$$
\noindent
The data for polarized SDM elements clearly rule out these  
predictions as is shown in Figs.~16 and 17 for $\pi^- p\to \pi^-  
\pi^+ n$ at 17.2 GeV/c. We find that $\rho^y_{ss} + \rho^y_{00} +  
2\rho^y_{11}$ and $-2 (\rho^y_{00} - \rho^y_{11})$ have large  
magnitudes but opposite signs while $2\rho^y_{1-1}$ has a small  
magnitude. The interference terms ${\rm Re} \rho^y_{10}$, ${\rm Re}  
\rho^y_{1s}$ and ${\rm Re} \rho^y_{0s}$ are all dissimilar and have  
large nonzero values. On the basis of this evidence we again must  
conclude that the past determinations of $\pi\pi$ phase shifts from  
unpolarized data on $\pi N \to \pi^+ \pi^- N$ are questionable.

The assumption of absence of $A_1$ exchange amplitudes means that  
pion production in $\pi N \to \pi^+ \pi^- N$ reactions does not  
depend on nucleon spin. What the measurements of $\pi N \to \pi^+  
\pi^- N$ on polarized targets found is that the pion production  
depends strongly on nucleon spin. The dynamics of the pion  
production is not as simple as has been assumed in the past  
determinations of $\pi\pi$ phase shifts. New determinations of  
$\pi\pi$ phase shifts are now required that do take into account the  
existence of $A_1$ exchange. Since the contributions of $A_1$  
exchange amplitudes are large and nontrivial, the revisions of  
$\pi\pi$ phase shifts will be significant. The new revised $S$-wave  
phase shift $\delta_0^0$ is then expected to show evidence for  
narrow scalar state $\sigma(750)$ in agreement with the measurements  
on polarized targets.
\bigskip
\noindent
{\bf VIII. Questions concerning evidence for narrow $\sigma(750)$}
\medskip
\noindent
{\bf A. Up-Down ambiguity and analyticity constraints.}
\medskip
Recently it has been claimed$^{49,50}$ that $\pi\pi$ phase shift  
$\delta^0_0$ can be determined from the $S$-wave intensities $I_S$  
obtained in our amplitude analysis of $\pi^- p \to \pi^- \pi^+ n$ on  
polarized target at 17.2 GeV/c, and that it would show the old  
Up-Down ambiguity of $\delta_0^0$. Only Up solution indicates a  
narrow $\sigma$ state and it is excluded because it is inconsistent  
with Roy equations.$^{51}$ From this it was concluded that  
$\sigma(750)$ does not exist$^{49}$ or that the evidence must be  
treated with reservation.$^{50}$

To answer this objection we first recall from (2.15) that
$$I_S = (|S_0|^2 + |S_1|^2)\Sigma\eqno(7.1)$$
\noindent
Here the amplitude $S_1$ is connected to $\delta^0_0$ through (7.1)  
and (7.3), and $S_0$ is the unknown $A_1$ exchange amplitude. It is  
obvious from this expression that the determination of $\delta_0^0$  
from data on $I_S$ depends on the model used for $A_1$ exchange  
amplitude  $S_0$. The data on polarized target require large $A_1$  
exchange amplitudes. At present the $A_1$ exchange amplitudes are  
not known. We must therefore conclude that the phase shift  
$\delta_0^0$ cannot be determined from the data on $S$-wave  
intensity $I_S$ at present.

Nevertheless, the data on $I_S$ do tell us something very important  
about the solutions for $\delta_0^0$. There are four solutions for  
$I_S$: $I_S(1,1),\ldots, I_S(2,2)$. Consequently there will be a  
fourfold ambiguity in $\delta_0^0$ for any given model of $A_1$  
exchange amplitude $S_0$. However, as can be seen in Fig.~10, the  
four solutions for $I_S$ are all very similar quantitatively.  
Consequently the four solutions for $\delta_0^0$ are expected to be  
very close to each other and similar. This contrasts with the large  
differences between the old Up and Down solutions. Fig.~18 shows  
$S$-wave intensity normalized to 1 at maximum for Down (curve A) and  
Up (curve B) solutions from the typical analysis of Estabrooks et  
al.$^{27,28}$ The large differences between the Up and Down  
solutions contrast sharply with the small differences shown between  
$S$-wave intensities $I_S(1,1)$ and $I_S(2,2)$ in Fig.~18. On the  
basis of the similar behaviour of all solutions for $I_S$ we do not  
anticipate the emergence of the old Up-Down ambiguity problem in  
$\delta_0^0$. It is even possible that the small differences between  
the four solutions for $I_S$ can be explained entirely as a small  
ambiguity in $A_1$ exchange amplitude $S_0$ leading to a unique  
determination of $\delta_0^0$ from the data on polarized target.

The above discussion applies also to the determination of $P$-wave  
phase shift $\delta_1^1$ from $I_L = (|L_0|^2 + |L_1|^2)\Sigma$. The  
amplitude $L_1$ is connected to $\delta_1^1$ by (7.1) and (7.3)  
while $L_0$ is another unknown $A_1$-exchange amplitude. The four  
solutions for $I_L$ are again very close so we expect similar  
solutions for $\delta_1^1$.

Assuming a model for $A_1$ exchange amplitudes $S_0$ and $L_0$, the  
obtained phase shifts $\delta_0^0$ and $\delta_1^1$ can be tested  
for consistency with dispersion relations$^{51}$ (Roy equations). If  
an inconsistency is found it means that we have to modify our model  
for $A_1$ exchange amplitudes $S_0$ and $L_0$, and try again. It is  
important to realize that Roy equations do not test the validity of  
the experimentally measured amplitudes $|S|^2$, $|\overline S |^2$,  
$|L|^2$, $|\overline L |^2$ or intensities $I_S$ and $I_L$. The Roy  
equations are constraints only on $\pi\pi$ phase shifts which  
follow from the analyticity properties of partial wave amplitudes in  
$\pi\pi \to \pi\pi$ scattering. However the requirement of  
consistency of phase shifts with the Roy equations can be used to  
constrain the possible models of $A_1$ exchange amplitudes.

We conclude that the experimental evidence for the narrow state  
$\sigma(750)$ is not in contradiction with analyticity and  
dispersion relations for $\pi\pi$ partial waves. The existence of  
$A_1$-exchange and narrow $\sigma(750)$ are experimental findings  
from measurements on polarized targets independent of the Roy  
equations. These experimental facts cannot be refuted by comparisons  
with standard phase shifts because these were obtained using an  
invalid assumption of absence of $A_1$-exchange.

\bigskip
\noindent
{\bf B. Absence of $\sigma(750)$ in $\gamma\gamma \to \pi^+ \pi^-$  
and central production $pp\to pp\pi^+ \pi^-$.}
\medskip
Morgan and Pennington suggested to discount the evidence for  
existence of narrow $\sigma(750)$ in $\pi N \to \pi^+ \pi^-N$  
because this state has not been observed in $\gamma\gamma \to \pi^+  
\pi^-$ reaction$^{49}$ and in central production$^{49, 52}$ $pp\to  
pp\pi^+ \pi^-$. However there are good reasons why one would not  
expect to observe narrow $\sigma(750)$ in these processes.

In the next section we shall argue that the narrow $\sigma(750)$ is  
the lowest mass scalar gluonium $0^{++} (gg)$. The principal  
support for this proposal is precisely the fact that $\sigma(750)$  
state is not observed in $\gamma\gamma \to \pi^+ \pi^-$ reaction.  
Since gluons do not couple directly to the photons, we expect  
$\sigma(750)$ not to appear in reaction $\gamma\gamma\to \pi^+  
\pi^-$ if it is pure gluonium or if it has only a small $q\overline  
q$ component.

The reaction $pp\to pp\pi^+ \pi^-$ was measured$^{53}$ at the CERN  
Intersecting Storage Rings (ISR) in a search for scalar gluonium.  
The structures reported in the moments $H(11)$ and $H(31)$ near  
$m(\pi^+ \pi^-) \approx 750$ MeV are consistent with $\sigma(750)$  
and $\rho^0(770)$ interference.

Assuming parity conservation there are 5 $S$-wave amplitudes and 15  
$P$-wave amplitudes in this reaction. The $\sigma(750)$ state may  
contribute only to some $S$-wave amplitudes and not to the others,  
as it does in $\pi^- p \to \pi^-\pi^+ n$  with amplitudes  
$|\overline S|^2\Sigma$ and $|S|^2\Sigma$. As we see in Fig.~10, the  
$S$-wave intensity $I_S(2,2)$ does not immediately suggest  
existence of narrow $\sigma(750)$. With 5 $S$-wave amplitudes in  
$pp\to pp\pi^+\pi^-$ it is very likely that $\sigma(750)$ stays  
hidden. We can observe $\sigma(750)$ in $\pi^- p \to \pi^- \pi^+ n$  
and $\pi^+ n \to \pi^+ \pi^- p$ reactions only when these production  
processes are measured on polarized targets, and the $S$- and  
$P$-wave amplitudes can be separated in a model independent way. For  
the same reasons we may see $\sigma(750)$ in central production  
$pp\to pp \pi^+ \pi^-$ only when measurements with polarized initial  
protons are made and the resonating $S$-wave amplitudes can be  
isolated. The ISR experiment does not separate the $S$- and $P$-wave  
amplitudes, and thus it is not conclusive.

\bigskip
\noindent
{\bf C. Comparison with other results for $\sigma$ state.}
\medskip
DM2 Collaboration measured$^{54}$ $\pi^+\pi^-$ mass distribution in  
$J/\psi \to \omega \pi^+ \pi^-$ decays and observed a quite broad  
low mass resonance (see Fig.~13a of Ref.~54). Interpreted as an  
$I=0$ $0^{++}$ $\sigma$ state, a single Breit-Wigner fit gives  
$m_\sigma = (414 \pm 20)$ MeV, $\Gamma_\sigma = (494 \pm 58)$ MeV.  
There is no indication for such state in our data on $S$-wave  
intensity $I_S$ in $\pi^+ n \to \pi^+ \pi^- p$ at 5.98 and 11.85  
GeV/c (see Fig.~12 and 13 above). The reasons for the discrepancy  
are not clear at the present.

Several recent theoretical analyses$^{55, 56, 57}$ claimed  
existence of a $\sigma$ meson with a mass around 1000 MeV and a  
broad width of 460--880 MeV. These analyses use as an input the  
$S$-wave phase shift $\delta^0_0$ and thus neglect the $A_1$  
exchange and other spin effects observed in pion production (see  
e.g. eq.~(5) in Ref.~56). It is possible that when these analyses  
include in their fits $A_1$ exchange that they will find a narrow  
$\sigma$ in agreement with the CERN data on polarized targets.

\bigskip
\noindent
{\bf IX. Constituent structure of the $\sigma(750)$ resonance.}
\medskip
In the usual quark model meson resonances are $q\overline q$  
states. The mass of $\sigma(750)$ is too low for it to be a  
$q\overline q$ state. The mass $M$ of the $q\overline q$ state  
increases with its angular momentum $L$ as $M = M_0 (2n+L)$ where  
$n$ is the degree of radial excitation. The lowest mass scalar  
mesons are ${}^3P_0$ states with masses expected to be around 1000  
MeV or higher.

It was suggested that $0^{++}(700)$ could be a four-quark  
$q\overline q q\overline q$ state in the MIT bag model.$^{58}$  
However, more detailed studies of $q\overline q q\overline q$  
systems conclude that pure multiquark hadrons do not exist$^{59,60}$  
with $\pi^+\pi^-$ decay.$^{61}$ We can also exclude the possibility  
that $\sigma(750)$ is a hybrid state $q\overline q g$. The lowest  
mass hybrid state must be a $0^{-+}$ or $1^{-+}$ state. Calculations  
based on bag models, QCD sum rules, lattice QCD and a string model  
all estimate$^{62}$ the masses of $0^{++} (q\overline q g)$ states  
to be above 1500 MeV.

Ellis and Lanik discussed the couplings of scalar gluonium $\sigma$  
on the basis of the low energy theorems of broken chiral symmetry  
and scale invariance, implemented using a phenomenological  
lagrangian.$^{63}$ They obtained for $\sigma \to \pi^+ \pi^-$ decay  
the following partial width
$$\Gamma (\sigma \to \pi^+ \pi^-) = {{(m_\sigma)^5}\over{48\pi  
G_0}}\eqno(9.1)$$
\noindent
where $G_0 \equiv <0|(\alpha_s/\pi)F_{\mu \nu} F^{\mu \nu} |0>$ is  
the gluon-condensate term$^{64}$ parametrizing the non-perturbative  
effects in QCD. The numerical values were estimated by the ITEP  
group$^{64}$ to be $G_0 \approx 0.012$ (GeV)$^4$ or up to $G_0  
\approx 0.030$ (GeV)$^4$ in later calculations.$^{65,66}$ It is very  
interesting to note, that when we take $G_0 = 0.015$ (GeV)$^4$ the  
Ellis-Lanik theorem (9.1) predicts partial width of $\sigma \to  
\pi^+\pi^-$ decay $\Gamma = 107$ MeV for the mass $m_\sigma = 753$  
MeV. This result is in perfect agreement with (4.20), the solution  
and method average values of mass and width of $\sigma(750)$ from  
the best fit to the measured mass distribution $|\overline S  
|^2\Sigma$ (Table 4). When we use for $m_\sigma$ the value 768 MeV  
obtained in interference fits with $f_0(980)$ then Ellis-Lanik  
theorem predicts a width $\Gamma(\sigma\to\pi^+\pi^-) = 118$ MeV,  
again in perfect agreement with (5.6) where $\Gamma_\sigma = 115 \pm  
38$ MeV. From this agreement we can conclude that the $\sigma(750)$  
is best understood as the lowest mass gluonium state $0^{++}(gg)$.

The gluonium interpretation of $\sigma(750)$ gathers further  
support from the lack of observation of $\sigma(750)$ in the  
reactions $\gamma\gamma \to\pi^+\pi^-$ and $\gamma\gamma\to\pi^0  
\pi^0$. Since gluons do not couple directly to photons we expect  
$\sigma(750)$ not to appear in reactions $\gamma\gamma\to\pi\pi$ if  
it is a pure gluonium state or if it contains only a small  
$q\overline q$ component. This conclusion is supported by the PLUTO  
and DELCO data.$^{67,68}$ However, the more recent DM1/2  
data$^{69,70}$ show an excess over the Born term expectation that is  
attributed to the formation of a broad scalar resonance with a  
two-photon width of $(10\pm 6)$ MeV. This would suggest some  
$q\overline q$ component in the $\sigma(750)$ state. The most recent  
results$^{71}$ are on $\gamma\gamma\to\pi^0\pi^0$ which show no  
evidence for a scalar state near 750 MeV.

Lattice QCD calculations by several groups$^{72-75}$ initially  
concluded that the gluonium ground state $0^{++}(gg)$ has a mass  
near the $\rho^0$ meson: 740$\pm$ 40 MeV. The most recent lattice  
QCD calculations predict much higher mass of the lowest scalar  
gluonium: the UKQCD group$^{76}$ predicts $1550\pm 50$ MeV while the  
IBM group$^{77,78}$ predicts $1740\pm 70$ MeV. However, it is  
important to remember that these calculations are for quenched QCD  
so that there is no coupling of the primitive gluonium to quarks.  
The coupling of gluonium to two pseudoscalars may have a significant  
effect on the gluonium mass and width.$^{52}$

We conclude that while the gluonium interpretation of the  
$\sigma(750)$ state is in agreement with low energy theorems of  
broken chiral symmetry and scale invariance, it is at variance with  
the most recent lattice QCD calculations. It is necessary to study  
this discrepancy and understand its origins and implications.

Finally we note that the anomalous energy dependence of $pp$ and  
$np$ elastic polarizations and the departure from the mirror  
symmetry in $\pi N$ elastic polarizations at intermediate energies  
require a low-lying Regge trajectory$^{79,80}$ corresponding to  
$\sigma(750)$. These anomalous structures in the polarization data  
may have been the first evidence for a gluonium exchange in two-body  
reactions.

\bigskip
\noindent
{\bf X. Amplitude spectroscopy -- a new direction in hadron  
spectroscopy.}
\medskip
The vast majority of hadron resonances have been identified through  
study of mass distributions of spin averaged cross-sections.  
Experiments with polarized targets opened a whole new approach to  
experimental hadron spectroscopy by making accessible the study of  
hadron production on the level of spin dependent production  
amplitudes. We may refer to this new approach to detecting and  
studying hadron resonances as amplitude spectroscopy.

This work represents the first effort to determine resonance  
parameters directly from a measured spin dependent production  
amplitude. In this case it was the amplitude $|\overline S  
|^2\Sigma$ measured in $\pi^- p_\uparrow \to \pi^-\pi^+ n$ at 17.2  
GeV/c. We have found that the best Breit-Wigner fits to the  
resonance $\sigma(750)$ on the amplitude level differ markedly from  
the Breit-Wigner fits to the spin-averaged $S$-wave intensities, and  
provide a more reliable information about the resonance parameters.

However, the significance of amplitude spectroscopy goes far beyond  
the more precise determinations of resonance parameters. The  
amplitude spectroscopy opens several prospects for new physics:
\medskip
\noindent
(a) Let us call dominant resonances those states which can be  
experimentally identified in spin-averaged cross-sections or Dalitz  
plots. Historically these dominant resonances led to the standard  
quark model and to QCD. However, a new species of subdominant  
resonances may exist which can be identified only at the level of  
spin dependent production amplitudes. The existence and properties  
of subdominant resonances could reveal new components of hadron  
structure to which unpolarized experiments are totally blind and  
could lead us beyond the standard quark model and standard QCD.
\medskip
\noindent
(b) The production of resonances (dominant and subdominant) may  
depend on nucleon spin and this dependence will provide important  
information about the dynamics of hadron interactions and about the  
very nature of hadron resonances.
\medskip
\noindent
(c) Standard QCD predicts new kinds of resonances such as dibaryon,  
gluonium and hybrid states. Many of these states may not be  
observable in the spin-averaged measurements which could explain the  
limitted success in identifying these states so far.
\medskip
The $\sigma(750)$ is the first example of a subdominant resonance  
observable only on the level of measured spin dependent production  
amplitudes. We have interpreted this resonance as the lowest mass  
gluonium $0^{++}(gg)$ in the Section IX. However, we must be open  
also to the possibility that $\sigma(750)$ represents the first  
signal of a new physics beyond the standard quark model and QCD. It  
may indicate the existence of a new component in hadron structure.

The measurements$^{18}$ of $K^+ n \to K^+ \pi^- p$ reactions at  
5.98 and 11.85 GeV/c at CERN-PS also allowed a model independent  
amplitude analysis$^{81}$ of this reaction at 5.98 GeV/c. The  
results of the amplitude analysis are in excellent agreement with  
Additive Quark Model predictions.$^{82}$ The data also  
suggest$^{18,81}$ the existence of another subdominant resonance,  
$I={1\over 2}\  0^{++}(890)$ with a narrow width of about 20 MeV.  
Such new resonance $\kappa(890)$ under $K^{0*}(892)$ could also  
signal new physics beyond the standard QCD. One such possibility is  
discussed in Ref.~81.

We recall that complete measurements of spin observables in  
two-body and quasi two-body reactions enable construction of spin  
amplitudes. The spin amplitudes show dip structures in their moduli  
associated with rapid and large changes of their relative phases.  
These dip structures resemble absorption resonances and systematic  
study of dip structures in two-body and other exclusive processes  
like $\pi N \to \pi^+ \pi^- N$ is a natural extension of hadron  
spectroscopy into the space-like region. Study of time-like  
resonances and space-like dips in spin dependent amplitudes should  
bring entirely new insights into the hadron dynamics and structure.

To explore these new frontiers of hadron spectroscopy and hadron  
dynamics, new advanced hadron facilities dedicated in large part to  
measurements with spin will be required. The proposed Canadian KAON  
Factory,$^{83}$ Los Alamos Hadron Facility$^{84}$ and European  
Hadron Facility$^{85}$ could integrate the new advanced technologies  
of polarized beams and targets in a single spin physics  
facility$^{86}$ to systematically advance the exploration and  
development of these new frontiers in hadron physics.$^{86-88}$

\bigskip
\noindent
{\bf XI. Summary}
\medskip
The measurements of reactions $\pi^-p_\uparrow \to \pi^- \pi^+n$ at  
17.2 GeV/c and $\pi^+ n_\uparrow \to \pi^+ \pi^- p$ at 5.98 and  
11.85 GeV/c on polarized target provide model-independent and  
solution-independent evidence for a narrow scalar state  
$\sigma(750)$. The amplitude analyses of $\pi^- p_\uparrow \to  
\pi^-\pi^+ n$ at small $t$ using $\chi^2$ minimization method$^{13}$  
and Monte Carlo method$^{23}$ yield very similar results for moduli  
of transversity amplitudes and cosines of their relative phases. In  
particular they agree that the transversity ``up'' $S$-wave  
amplitude $|\overline S |^2\Sigma$ resonantes near 750 MeV while the  
transversity ``down'' amplitude $|S|^2\Sigma$ is nonresonating and  
constitutes a large background in the spin-averaged $S$-wave  
intensity $I_S = (|S|^2 + |\overline S |^2)\Sigma$. For this reason  
it is preferable to determine resonance parameters of $\sigma(750)$  
directly from the measured mass distribution of $|\overline S  
|^2\Sigma$.

We have performed several types of Breit-Wigner fits to $|\overline  
S |^2\Sigma$. We have shown that the Pi\v s\'ut-Roos resonance  
shape formula and phenomenological shape formula give similar  
results. Single Breit-Wigner fits yield a width of $\sigma(750)$ in  
the range 192--256 MeV. We have studied the effect of background in  
three approaches: incoherent background, constant coherent  
background, and $t$-averaged constant coherent background. The last  
method yields the best fit with the lowest $\chi^2$/dof. The  
solution and method average for the $\sigma$ mass and width from  
this best fit are
$$m_\sigma = 753 \pm 19\ \hbox{MeV}\quad ,\quad \Gamma_\sigma = 108  
\pm 53\ \hbox{MeV}\eqno(11.1)$$

We also performed the conventional fits to spin-averaged $S$-wave  
intensity $I_S$. We found again that the inclusion of background  
(incoherent in this case) reduces the fitted value of the $\sigma$  
width and improves $\chi^2$/dof. Nevertheless, the direct fits to  
$|\overline S |^2\Sigma$ are preferable at 17.2 GeV. Due to lower  
statistics at 5.98 and 11.85 GeV/c, we must use results for $I_S$ to  
obtain $\sigma$ resonance parameters. All four solutions resonate  
at these larger momentum transfers but yield a broader $\sigma$  
width: $\Gamma_\sigma = 195 \pm 81$ MeV at 5.98 GeV/c and  
$\Gamma_\sigma = 166 \pm 54$ MeV at 11.85 GeV/c.

We conclude that the best overall estimate of the mass and width of  
$\sigma(750)$ are the values in (11.1) from the best fit to  
$|\overline S|^2\Sigma$ (Table 4).

We have also examined the interference of $\sigma(750)$ with  
$f_0(980)$ and found that it has only a small effect on the mass and  
width of $\sigma(750)$. A fit to amplitude $|\overline S|^2\Sigma$  
in the mass range above 1120 MeV shows evidence for a scalar state  
with average mass $1280 \pm 12$ MeV and width $192 \pm 26$ MeV.

The conventional $S$-wave phase shifts $\delta_0^0$ show no  
evidence for the narrow $\sigma(750)$ state. It must be reiterated,  
that the past determinations of $\pi\pi$ phase shifts from  
unpolarized data on $\pi^- p \to \pi^- \pi^+ n$ assumed the absence  
of $A_1$-exchange amplitudes. This assumption is invalidated by  
measurements of $\pi^- p \to \pi^-\pi^+ n$, $\pi^+ n \to \pi^+\pi^-  
p$ and $K^+ n \to K^+ \pi^-p$ on polarized targets which find large  
and nontrivial $A_1$-exchange contributions. New determinations of  
$\pi\pi$ phase shifts are required that do take into account the  
existence of $A_1$-exchange. Since $A_1$-exchange contributions are  
large, the revisions of $\pi\pi$ phase shifts will be significant  
and should provide evidence for a narrow $\sigma(750)$ state in  
agreement with the CERN data on polarized targets.

The mass of $\sigma(750)$ is too low for it to be a $q\overline q$  
state. We proposed to identify $\sigma(750)$ with the lowest mass  
scalar gluonium $0^{++} (gg)$. This proposal is supported by the  
perfect agreement with the Ellis-Lanik theorem (9.1) relating the  
decay width of scalar gluonium $\Gamma (\sigma\to \pi^+\pi^-)$ to  
its mass $m_\sigma$. Another experimental support for the gluonium  
interpretation of $\sigma(750)$ is its absence in $\gamma\gamma \to  
\pi^+\pi^-$ reaction. However, the low mass of $\sigma(750)$ is at  
variance with the more recent calculations of lattice QCD which  
predict masses of scalar gluonium above 1500 MeV.

Experiments with polarized targets have opened a whole new approach  
to experimental hadron spectroscopy by making accessible the study  
of hadron production on the level of production spin amplitudes. We  
may expect that this new field of amplitude spectroscopy will be  
further developed at the new proposed advanced hadron  
facilities.$^{83-88}$
\bigskip
\bigskip
\noindent
{\bf Acknowledgements.}
\medskip
I wish to thank M.D.~Scadron, J.~Stern and especially  
N.A.~T\"ornqvist for their interest and stimulating e-mail  
correspondence. This work was supported by Fonds pour la Formation  
de Chercheurs et l'Aide \`a la Recherche (FCAR), Minist\`ere de  
l'Education du Qu\`ebec, Canada.

\bigskip
\noindent
\vfill\eject
\noindent
{\bf References.}
\medskip
\item{1.} K.R.~Popper, the Logic of Scientific Discovery, New York,  
Basic Books 1959 and Harper Torchbooks, 1968.
\smallskip
\item{2.} A.~de Lesquen {\it et al.}, Phys.~Lett. \underbar{40B},  
277 (1972).
\smallskip
\item{3.} G.~Cozzika {\it et al.}, Phys.~Lett. \underbar{40B}, 281  
(1972).
\smallskip
\item{4.} M.~Fujisaki {\it et al.}, Nucl.~Phys. \underbar{B152},  
232 (1979).
\smallskip
\item{5.} A.~Yokosawa, Phys.~Rep. \underbar{64}, 50 (1980).
\smallskip
\item{6.} I.P.~Aur {\it et al.}, Phys.~Rev. \underbar{D32}, 1609 (1985).
\smallskip
\item{7.} A.~Rahbar {\it et al.}, Phys.~Rev.~Lett. \underbar{47},  
1811 (1981).
\smallskip
\item{8.} Spin Observables of Nuclear Probes, Telluride 1988,  
Editor Ch.~J.~Horowitz.
\smallskip
\item{9.} M.~Svec, Phys.~Rev. \underbar{D22}, 70 (1980).
\smallskip
\item{10.} G.~Lutz and K.~Rybicki, Max Planck Institute, Munich,  
Internal Report No. MPI-PAE/Exp.~E1.75, 1978 (unpublished).
\smallskip
\item{11.} G.~Grayer {\it et al.}, Nucl.~Phys. \underbar{B75}, 189  
(1974).
\smallskip
\item{12.} J.G.H.~de Groot, PhD Thesis, University of Amsterdam,  
1978 (unpublished).
\smallskip
\item{13.} H.~Becker {\it et al.}, Nucl.~Phys. \underbar{B150}, 301  
(1979).
\smallskip
\item{14.} H.~Becker {\it et al.}, Nucl.~Phys. \underbar{B151}, 46  
(1979).
\smallskip
\item{15.} V.~Chabaud {\it et al.}, Nucl.~Phys. \underbar{B223}, 1  
(1983).
\smallskip
\item{16.} K.~Rybicki, I.~Sakrejda, Z.~Phys. \underbar{C28}, 65 (1985).
\smallskip
\item{17.} A.~de Lesquen {\it et al.}, Phys.~Rev. \underbar{D32},  
21 (1985).
\smallskip
\item{18.} A.~de Lesquen {\it et al.}, Phys.~Rev. \underbar{D39},  
21 (1989).
\smallskip
\item{19.} M.~Svec, A.~de Lesquen, L.~van Rossum, Phys.~Rev.  
\underbar{D42} 934 (1990).
\smallskip
\item{20.} M.~Svec, A.~de Lesquen, L.~van Rossum, Phys.~Rev.  
\underbar{D45} 55 (1992).
\smallskip
\item{21.} M.~Svec, A.~de Lesquen, L.~van Rossum, Phys.~Rev.  
\underbar{D45} 1518 (1992).
\smallskip
\item{22.} M.~Svec, A.~de Lesquen, L.~van Rossum, Phys.~Rev.  
\underbar{D46} 946 (1992).
\smallskip
\item{23.} M.~Svec, Phys.~Rev. \underbar{D53}, 2343 (1996).
\smallskip
\item{24.} S.D.~Protopopescu {\it et al.}, Phys.~Rev.  
\underbar{D7}, 1279 (1973).
\smallskip
\item{25.} W.~Ochs, PhD Thesis, Ludwig-Maximilians-Universit\"at,  
M\"unchen, 1973.
\smallskip
\item{26.} B.~Hyams {\it et al.}, Nucl.~Phys. \underbar{B64}, 134 (1973).
\smallskip
\item{27.} P.~Estabrooks, A.D.~Martin, $\pi\pi$ Scattering-1973,  
Proceedings of the Int. Conf. on $\pi\pi$ Scattering, Tallahassee,  
1973, edited by D.K.~Williams and V.~Hagopian, AIP Conf. Proc. No.  
13 (AIP, New York, 1973), p.~37.
\smallskip
\item{28.} P.~Estabrooks and A.D.~Martin, Nucl.~Phys.  
\underbar{B79}, 301 (1974).
\smallskip
\item{29.} P.~Estabrooks and A.D.~Martin, Nucl.~Phys.  
\underbar{B95}, 322 (1975).
\smallskip
\item{30.} K.L.~Au, D.~Morgan, M.R.~Pennington, Phys.~Rev.  
\underbar{D35}, 1633 (1987).
\smallskip
\item{31.} B.S.~Zou and D.V.~Bugg, Phys.~Rev. \underbar{D48}, R3948  
(1993).
\smallskip
\item{32.} M.~Anselmino, A.~Efremov, E.~Leader, Phys.~Rep.  
\underbar{261}, 1 (1995).
\smallskip
\item{33.} J.T.~Donohue and Y.~Leroyer, Nucl.~Phys.  
\underbar{B158}, 123 (1979).
\smallskip
\item{34.} V.P.~Kanavets et al., in Workshop on High Energy Spin  
Physics, Protvino 1995, Ed. S.B.~Nurushev.
\smallskip
\item{35.} C.D.~Lac {\it et al.}, J.~Phys. (France) \underbar{51},  
2689 (1990).
\smallskip
\item{36.} P.W.~Johnson, R.C.~Miller and G.M.~Thomas, Phys.~Rev.  
\underbar{D7}, 1895 (1977).
\smallskip
\item{37.} H.~Palka, Institute of Nuclear Physics, Cracow, Internal  
Report No. 1230/PH, 1983, Appendix D (unpublished).
\smallskip
\item{38.} F.~James, in Techniques and Concepts of High Energy  
Physics II, edited by Thomas Ferbel (Plenum, New York, 1983),  
p.~196.
\smallskip
\item{39.} J.~Pi\v s\'ut and M.~Roos, Nucl.~Phys. \underbar{B6}  
(1968) 325.
\smallskip
\item{40.} M.~Aguilar-Bem\'itez and J.A.~Rubio, Hadronic Resonances  
Part II: Experimental Analyses, published by Grupo  
Inter-universitario de Fisica Teorica de Atlas Energias as Report  
GIFT4/75. Vol.~2, 1975, Facultad de Ciencias, Universidad de  
Zaragoza, Spain.
\smallskip
\item{41.} H.~Pilkuhn, The Interactions of Hadrons, North-Holland  
Publishing Company, 1967.
\smallskip
\item{42.} E~Byckling and K.~Kajantie, Particle Kinematics, Wiley, 1973.
\smallskip
\item{43.} S.~Humble, Introduction to Particle Production in Hadron  
Physics, Academic Press 1974.
\medskip
\item{44.} B.R.~Martin, D.~Morgan, G.~Shaw, Pion-Pion Interactions  
in Particle Physics, Academic Press 1976.
\smallskip
\item{45.} I.~Silin, computer code FUMILI, Long Write-up D510, CERN  
Computer Centre Program Library, revised 1985.
\smallskip
\item{46.} J.M.~Blatt and V.F.~Weiskopf, Theoretical Nuclear  
Physics, Wiley, New York, 1952.
\smallskip
\item{47.} N.A.~T\"ornqvist, private e-mail correspondence.
\smallskip
\item{48.} W.D.~Apel et al., Phys.~Lett. \underbar{41B}, 542 (1972).
\smallskip
\item{49.} D.~Morgan, in Hadron 93, Como, Italy, 1993, Nuov.~Cim.  
\underbar{107A}, 1883 (1994).
\smallskip
\item{50.} Review of Particle Properties, Phys.~Rev.  
\underbar{D50}, 1478 (1994).
\smallskip
\item{51.} M.R.~Pennington, S.D.~Protopopescu, Phys.~Rev.  
\underbar{D7}, 2591 (1973).
\smallskip
\item{52.} M.R.~Pennington, in Hadron 95, Manchester 1995, Editors  
M.K.~Birke, G.~Lafferty, J.~McGovern, World Scientific 1996, p.~3.
\smallskip
\item{53.} T.~Akesson et al., Phys.~Lett. \underbar{133B}, 263 (1983).
\smallskip
\item{54.} DM2 Collaboration, J.~Augustin et al., Nucl.~Phys.  
\underbar{B320}, 1 (1989).
\smallskip
\item{55.} N.N.~Achasov, G.N.~Shestakov, Phys.~Rev. \underbar{D49},  
5779 (1994).
\smallskip
\item{56.} V.V.~Anisovich et al., Phys.~Lett. \underbar{B355}, 363  
(1995).
\smallskip
\item{57.} N.A.~T\"ornqvist, M.~Roos, Phys.~Rev.~Lett.  
\underbar{76}, 1575 (1996).
\smallskip
\item{58.} R.J.~Jaffe, Phys.~Rev. \underbar{D15}, 267 (1977).
\smallskip
\item{59.} R.P.~Bickerstaff, B.H.~McKellar, Z.~Phys.  
\underbar{C16}, 171 (1982).
\smallskip
\item{60.} J.~Weinstein and N.~Isgur, Phys.~Rev. \underbar{D27},  
588 (1983).
\smallskip
\item{61.} B.A.~Liu and K.F.~Liu, Phys.~Rev. \underbar{D30}, 613 91984).
\smallskip
\item{62.} F.E.~Close, Nucl.~Phys. \underbar{A416}, 55c (1984).
\smallskip
\item{63.} J.~Ellis and J.~Lanik, Phys.~Lett. \underbar{150B}, 289  
(1985).
\smallskip
\item{64.} M.S.~Shifman, A.I.~Vainshtein, V.I.~Zakharov,  
Nucl.~Phys. \underbar{B147}, 385 (1975); ibid \underbar{B147}, 448  
(1979).
\smallskip
\item{65.} J.S.~Bell and R.A.~Bertlmann, Nucl.~Phys.  
\underbar{B177}, 218 (1981); ibid \underbar{B187}, 285 (1981).
\smallskip
\item{66.} A.~Bradley, C.S.~Largensiepen and G.~Shaw, Phys.~Lett.  
\underbar{102B}, 359 (1981).
\smallskip
\item{67.} Ch.~Berger et al., Z.~Phys. \underbar{C26}, 199 (1984).
\smallskip
\item{68.} H.~Aihara et al., Phys.~Rev.~Lett. \underbar{57}, 404 (1986).
\smallskip
\item{69.} A.~Coureau et al., Nucl.~Phys. \underbar{B271}, 1 (1986).
\smallskip
\item{70.} Z.~Ajaltouni et al., Phys.~Lett. \underbar{194B}, 573  
(1987); \underbar{197B}, 565E (1987).
\smallskip
\item{71.} H.~Marsiske et al., Phys.~Rev. \underbar{D41}, 3324 (1990).
\smallskip
\item{72.} K.~Ishikawa et al., Phys.~Lett. \underbar{116B}, 429 (1982).
\smallskip
\item{73.} K.~Ishikawa et al., Z.~Phys.~C \underbar{21}, 327  
(1983); \underbar{21}, 167 (1983).
\smallskip
\item{74.} B.~Berg and A.~Billoire, Nucl.~Phys. \underbar{B221},  
109 (1983).
\smallskip
\item{75.} W.H.~Hamber and M.~Urs Meiller, Phys.~Rev.  
\underbar{D29}, 928 (1984).
\smallskip
\item{76.} G.~Bali et al., Phys.~Lett. \underbar{B309}, 378 (1993).
\smallskip
\item{77.} F.~Butler et al., Phys.~Rev.~Lett. \underbar{70}, 2849 (1993).
\smallskip
\item{78.} M.~Chen et al., Nucl.~Phys. (Suppl.) \underbar{34}, 357  
(1994).
\smallskip
\item{79.} J.W.~Dash and H.~Navelet, Phys.~Rev. \underbar{D13},  
1940 (1976).
\smallskip
\item{80.} G.~Girardi and H.~Navelet, Phys.~Rev. \underbar{D14},  
280 (1976).
\smallskip
\item{81.} M.~Svec, A.~de Lesquen, L.~van Rossum, Phys.~Rev.  
\underbar{D45}, 1518 (1992).
\smallskip
\item{82.} M.~Svec, A.~de Lesquen, L.~van Rossum, Phys.~Lett.  
\underbar{220B}, 653 (1989).
\smallskip
\item{83.} Canadian KAON Factory (TRIUMF, Vancouver, 1985).
\smallskip
\item{84.} Physics and a Plan for a 45 GeV Facility [LANL Report  
No.~LA--10720--MS, Los Alamos 1986 (unpublished)].
\smallskip
\item{85.} Proceedings of the International Conference on a  
European Hadron Facility, Mainz, Germany, 1986, edited by T.~Walcher  
[Nucl.~Phys. \underbar{B279}, 2 (1987)].
\smallskip
\item{86.} M.~Svec, in Proceedings of the 8th International  
Symposium on High-Energy Spin Physics, Minneapolis, 1988, edited by  
K.J.~Heller, AIP Conf.~Proc.~No.~187 (AIP, New York, 1989), Vol.~2,  
p.~1181.
\smallskip
\item{87.} J.R.~Comfort, in Proceedings of a Workshop on Science at  
the KAON Factory, TRIUMF, 1990, edited by D.R.~Gill (unpublished),  
Vol.~2.
\smallskip
\item{88.} M.~Svec, in Proceedings of a Workshop on Future  
Directions in Particle and Nuclear Physics at Multi-GeV Hadron Beam  
Facilities, BNL, 1993, edited by D.F.~Geesaman (BNL Report  
No.~BNL--52389), p.~401.
\vfill\eject\

\tabskip=1em plus2em minus.5em
\halign to \hsize{\hfil # \hfil & \hfil #\hfil & \hfil # \hfil  
&\hfil #\hfil & \hfil # \hfil\cr
$|\overline  
S|^2\Sigma$&$m_\sigma$&$\Gamma_\sigma$&$N_\sigma$&$\chi^2$/dof\cr
Solution&(MeV)&(MeV)\cr
\noalign{\medskip\hrule\medskip}
\multispan3{Pi\v s\'ut-Roos shape formula\hfil}\cr
\noalign{\medskip}
1(MC) & 736 $\pm$ 6 & 230 $\pm$ 32 & 1.40 $\pm$ 0.12 & 0.388\cr
\noalign{\smallskip}
2(MC) & 745 $\pm$ 12 & 240 $\pm$ 59 & 1.71 $\pm$ 0.23 & 0.276\cr
\noalign{\medskip}
1$(\chi^2)$ & 738 $\pm$ 4 & 191 $\pm$ 16 & 1.50 $\pm$ 0.11 & 0.662\cr
\noalign{\smallskip}
2$(\chi^2)$ & 752 $\pm$ 10 & 253 $\pm$ 46 & 1.79 $\pm$ 0.15 & 0.968\cr
\noalign{\bigskip}
\multispan3{Phenomenological shape formula\hfil}\cr
\noalign{\medskip}
1(MC) & 732 $\pm$ 6 & 231 $\pm$ 33 & 6.50 $\pm$ 0.57 & 0.418\cr
\noalign{\smallskip}
2(MC) & 740 $\pm$ 11 & 241 $\pm$ 61 & 7.94 $\pm$ 1.11 & 0.288\cr
\noalign{\medskip}
1$(\chi^2)$ & 733 $\pm$ 4 & 192 $\pm$ 16 & 7.00 $\pm$ 0.50 & 0.740\cr
\noalign{\smallskip}
2$(\chi^2)$ & 747 $\pm$ 10 & 256 $\pm$ 47 & 8.29 $\pm$ 0.74 & 0.986\cr}
\bigskip
\bigskip
\noindent
{\bf Table 1.} Results of the fits to the mass distribution  
$|\overline S |^2\Sigma$ measured in $\pi^-  p \to \pi^- \pi^+ n$ at  
17.2 GeV/c using a single Breit-Wigner formula (4.1). The notation  
MC and $\chi^2$ indicates the solutions obtained by the Monte Carlo  
and $\chi^2$ minimization methods, respectively.
\vfill\eject

\halign to \hsize{#\hfil & \hfil #\hfil & \hfil #\hfil & \hfil  
#\hfil & \hfil #\hfil & \hfil #\hfil\cr
$|\overline S|^2\Sigma$ & $m_\sigma$ & $\Gamma_\sigma$ & $B$ &  
$N_S$ & $\chi^2$/dof\cr
Solution & (MeV) & (MeV)\cr
\noalign{\medskip\hrule\medskip}
\multispan3{Phenomenological shape formula\hfil}\cr
\noalign{\medskip}
1(MC) & 731 $\pm$ 6 & 202 $\pm$ 110 & 0.15 $\pm$ 0.55 & 5.75 $\pm$  
2.54 & 0.416\cr
\noalign{\medskip}
2(MC) & 744$\pm$14 &  103 $\pm$ 74 & 0.73 $\pm$ 0.49 & 5.11 $\pm$  
1.84 & 0.144\cr
\noalign{\medskip}
1$(\chi^2$) & 736 $\pm$ 4 & 147 $\pm$ 43 & 0.19 $\pm$ 0.17 & 6.13  
$\pm$ 0.84 & 0.696\cr
\noalign{\medskip}
2$(\chi^2)$ & 745 $\pm$ 41 & 98 $\pm$ 41 & 0.70 $\pm$ 0.28 & 5.70  
$\pm$ 1.30 & 0.626\cr}
\bigskip
\bigskip
\noindent
{\bf Table 2.} Results of the fits to the mass distribution  
$|\overline S |^2\Sigma$ measured in $\pi^- p \to \pi^- \pi^+ n$ at  
17.2 GeV/c using a Breit-Wigner formula with a constant incoherent  
background (4.6). The notation MC and $\chi^2$ as in Table 1.
\vfill\eject

\halign to \hsize{\hfil #\hfil & \hfil # \hfil & \hfil # \hfil &  
\hfil # \hfil & \hfil #\hfil & \hfil #\hfil & \hfil #\hfil\cr
$|\overline S |^2\Sigma$ & $m_\sigma$ & $\Gamma_\sigma$ & $B_1$ &  
$B_2$ & $N_S$ & $\chi^2$/dof\cr
Solution & (MeV) & (MeV)\cr
\noalign{\medskip\hrule\medskip}
\multispan3{Phenomenological shape formula\hfil}\cr
\noalign{\medskip}
1(MC) & 770 $\pm$ 19 & 114 $\pm$ 17 & 0.90 $\pm$ 0.43 & 1.34 $\pm$  
0.84 & 1.09 $\pm$ 0.61 & 0.136\cr
\noalign{\medskip}
2(MC) & 745 $\pm$ 31 & 104 $\pm$ 76 & 0.02 $\pm$ 1.07 & 1.84 $\pm$  
1.14 & 1.09 $\pm$ 0.90 & 0.144\cr
\noalign{\medskip}
1$(\chi^2)$ & 761 $\pm$ 13 & 138 $\pm$ 19 & 0.34 $\pm$ 0.16 & 0.69  
$\pm$ .045 & 2.41 $\pm$ 1.14 & 0.362\cr
\noalign{\medskip}
2$(\chi^2)$ & 738 $\pm$ 20 & 103 $\pm$ 112 & $-0.17 \pm 0.65$ &  
1.30 $\pm$ 0.50 & 1.79 $\pm$ 0.86 & 0.898\cr}
\bigskip
\bigskip
\noindent
{\bf Table 3.} Results of the fits to the mass distribution  
$|\overline S |^2\Sigma$ measured in $\pi^- p \to \pi^-\pi^+ n$ at  
17.2 GeV/c using a Breit-Wigner formula with constant coherent  
background (4.15). The notation MC and $\chi^2$ as in Table 1.
\vfill\eject

\halign to \hsize{\hfil #\hfil & \hfil # \hfil & \hfil # \hfil &  
\hfil # \hfil & \hfil #\hfil & \hfil #\hfil & \hfil #\hfil\cr
$|\overline S |^2\Sigma$ & $m_\sigma$ & $\Gamma_\sigma$ & $B$ & $B$  
& $N_S$ & $\chi^2$/dof\cr
Solution & (MeV) & (MeV)\cr
\noalign{\medskip\hrule\medskip}
\multispan 3{Phenomenological shape formula}\cr
\noalign{\medskip}
1(MC) & 774 $\pm$ 14 & 101 $\pm$ 44 & 0.73 $\pm$ 0.31 & 0.29 $\pm$  
0.14 & 3.99 $\pm$ 0.96 & 0.108\cr
\noalign{\medskip}
2(MC) & 744 $\pm$ 31 & 103 $\pm$ 79 & 0.73 $\pm$ 0.54 & 0.01 $\pm$  
0.23 & 5.10 $\pm$ 1.90 & 0.144\cr
\noalign{\medskip}
1$(\chi^2)$ & 761 $\pm$ 12 & 134 $\pm$ 41 & 0.25 $\pm$ 0.17 & 0.15  
$\pm$ 0.07 & 5.74 $\pm$ 0.82 & 0.362\cr
\noalign{\medskip}
2$(\chi^2)$ & 733 $\pm$ 20 & 93 $\pm$ 48 & $0.80 \pm 0.39$ & -0.12  
$\pm$ 0.19 & 5.31 $\pm$ 1.53 & 0.592\cr}
\bigskip
\bigskip
\noindent
{\bf Table 4.} Results of the fits to the mass distribution  
$|\overline S |^2\Sigma$ measured in $\pi^- p \to \pi^-\pi^+ n$ at  
17.2 GeV/c using a Breit-Wigner formula with $t$-averaged constant  
coherent background (4.17). The notation MC and $\chi^2$ as in Table  
1.
\vfill\eject

\halign to \hsize{\hfil #\hfil & \hfil # \hfil & \hfil # \hfil &  
\hfil # \hfil & \hfil #\hfil & \hfil #\hfil & \hfil #\hfil\cr
$|\overline S |^2\Sigma$ & $m_\sigma$ & $\Gamma_\sigma$ & $B_1$ &  
$B_2$ & $N_S$ & $\chi^2$/dof\cr
Solution & (MeV) & (MeV)\cr
\noalign{\medskip\hrule\medskip}
\multispan 3{Phenomenological shape formula}\cr
\noalign{\medskip}
1(MC, $\chi^2$) & 778 $\pm$ 13 & 95 $\pm$ 27 & 1.20 $\pm$ 0.35 &  
0.85 $\pm$ 0.33 & 1.24 $\pm$ 0.39 & 0.096\cr
\noalign{\medskip}
2(MC, $\chi^2$) & 758 $\pm$ 32 & 135 $\pm$ 49 & 0.54 $\pm$ 0.69 &  
1.00 $\pm$ 0.28 & 1.85 $\pm$ 0.83 & 0.162\cr
\noalign{\bigskip}
&&&$C_1$ & $C_2$\cr
\noalign{\medskip}
1(MC, $\chi^2$) &&& 0.42 $\pm$ 0.52 & 1.25 $\pm$ 0.39\cr
\noalign{\medskip}
2(MC, $\chi^2$) &&& $-0.35 \pm 0.67$ & 0.97 $\pm$ 0.55\cr}
\bigskip
\bigskip
\noindent
{\bf Table 5.} Results of the fit to the mass distribution  
$|\overline S|^2\Sigma$ in the mass range from 600 to 1120 MeV  
taking into account the interference of $\sigma(750)$ with  
$f_0(980)$ using the parametrization (5.3). The notation MC and  
$\chi^2$ as in Table 1.
\bigskip
\bigskip
\bigskip
\bigskip
\bigskip
\bigskip
\halign to \hsize{\hfil #\hfil & \hfil #\hfil & \hfil #\hfil &  
\hfil # \hfil & \hfil # \hfil & \hfil #\hfil\cr
$|\overline S|^2\Sigma$ & $m$ & $\Gamma$ & $B$ & $N$ & $\chi^2$/dof\cr
Solution & (MeV) & (MeV)\cr
\noalign{\medskip\hrule\medskip}
1($\chi^2$) & 1284 $\pm$ 12 & 209 $\pm$ 29 & 0.001 $\pm$ 0.32 &  
5.96 $\pm$ 0.62 & 1.393\cr
2($\chi^2$) & 1276 $\pm$ 11 & 175 $\pm$ 24 & 0.001 $\pm$ 0.09 &  
6.21 $\pm$ 0.70 & 1.738\cr}
\bigskip
\bigskip
\noindent
{\bf Table 6.} The results of the fit to the mass distribution  
$|\overline S |^2\Sigma$ in the $f_0 (1300)$ mass region from 1120  
to 1520 MeV using a single Breit-Wigner formula with incoherent  
constant background. The notation $\chi^2$ as in Table 1.
\vfill\eject

\halign to \hsize{\hfil #\hfil & \hfil #\hfil & \hfil #\hfil &  
\hfil # \hfil & \hfil # \hfil & \hfil #\hfil\cr
$I_S$ & $m_\sigma$ & $\Gamma_\sigma$ & $B$ & $N_S$ & $\chi^2$/dof\cr
Solution & (MeV) & (MeV)\cr
\noalign{\medskip\hrule\medskip}
\multispan3{Single Breit-Wigner fit\hfil}\cr
\noalign{\medskip}
(1,1) & 766 $\pm$ 5 & 258 $\pm$ 19 & -- & 1.98 $\pm$ 0.07 & 0.450\cr
\noalign{\medskip}
(1,2) & 769 $\pm$ 12 & 263 $\pm$ 45 & -- & 2.26 $\pm$ 0.17 & 0.498\cr
\noalign{\medskip}
(2,1) & 766 $\pm$ 10 & 255 $\pm$ 37 & -- & 2.19 $\pm$ 0.15 & 0.240\cr
\noalign{\medskip}
(2,2) & 768 $\pm$ 12 & 303 $\pm$ 49 & -- & 2.48 $\pm$ 0.16  & 0.816\cr
\noalign{\medskip}
\multispan5{Breit-Wigner fit with constant background\hfil}\cr
\noalign{\medskip}
(1,1) & 767$\pm$5 & 210 $\pm$ 43 & 0.19 $\pm$ 0.17 & 1.69 $\pm$  
0.23 & 0.365\cr
\noalign{\medskip}
(1,2) & 768 $\pm$ 12 & 209 $\pm$ 99 & 0.20 $\pm$ 40 & 1.92 $\pm$  
0.61 & 0.470\cr
\noalign{\medskip}
(2,1) & 766 $\pm$ 9 & 208 $\pm$ 82 & 0.17 $\pm$ 0.32 & 1.91 $\pm$  
0.47 & 0.218\cr
\noalign{\medskip}
(2,2) & 765 $\pm$ 10 & 188 $\pm$ 76 & 0.41 $\pm$ 0.34 & 1.85 $\pm$  
0.42 & 0.700\cr}
\bigskip
\bigskip
\noindent
{\bf Table 7.} Results of the fits to the four solutions of the  
$S$-wave intensity measured in $\pi^- p \to \pi^- \pi^+ n$ at 17.2  
GeV/c using Monte Carlo method for amplitude analysis. The fits are  
made with Breit-Wigner parametrization (6.3) and (6.4) with the Pi\v  
s\' ut-Roos shape factor.
\vfill\eject

\halign to \hsize{\hfil #\hfil & \hfil #\hfil & \hfil #\hfil &  
\hfil # \hfil & \hfil # \hfil & \hfil #\hfil\cr
$I_S$ & $m_\sigma$ & $\Gamma_\sigma$ & $B$ & $N_S$ & $\chi^2$/dof\cr
Solution & (MeV) & (MeV)\cr
\noalign{\medskip\hrule\medskip}
\multispan3{Single Breit-Wigner fit\hfil}\cr
\noalign{\medskip}
(1,1) & 760 $\pm$ 8 & 269 $\pm$ 29 & -- & 2.00 $\pm$ 0.13 & 0.414\cr
\noalign{\medskip}
(2,2) & 786 $\pm$ 21 & 408 $\pm$ 90 & -- & 2.24 $\pm$ 0.16 & 1.140\cr
\noalign{\medskip}
\multispan5{Breit-Wigner fit with constant background\hfil}\cr
\noalign{\medskip}
(1,1) & 761$\pm$8 & 227 $\pm$ 68 & 0.12 $\pm$ 0.19 & 1.84 $\pm$  
0.28 & 0.394\cr
\noalign{\medskip}
(2,2) & 780 $\pm$ 13 & 187 $\pm$ 77 & 0.63 $\pm$ 0.31 & 1.51 $\pm$  
0.31 & 0.864\cr}
\bigskip
\bigskip
\noindent
{\bf Table 8.} Results of the fits to two of four solutions of the  
$S$-wave intensity measured in $\pi^- p \to \pi^- \pi^+ n$ at 17.2  
GeV/c using $\chi^2$ minimization method for amplitude analysis. The  
fits are made with Breit-Wigner parametrization (6.3) and (6.4)  
with the Pi\v s\' ut-Roos shape factor.
\vfill\eject

\halign to \hsize{\hfil #\hfil & \hfil #\hfil & \hfil #\hfil &  
\hfil # \hfil & \hfil # \hfil & \hfil #\hfil\cr
$I_S$ & $m_\sigma$ & $\Gamma_\sigma$ & $B$ & $N_S$ & $\chi^2$/dof\cr
Solution & (MeV) & (MeV)\cr
\noalign{\medskip\hrule\medskip}
\multispan3{Single Breit-Wigner fit\hfil}\cr
\noalign{\medskip}
(1,1) & 723 $\pm$ 22 & 282 $\pm$ 68 & -- & 0.53 $\pm$ 0.10 & 0.888\cr
\noalign{\medskip}
(1,2) & 696 $\pm$ 36 & 333 $\pm$ 128 & -- & 1.13 $\pm$ 0.34 & 0.118\cr
\noalign{\medskip}
(2,1) & 740 $\pm$ 32 & 296 $\pm$ 116 & -- & 1.02 $\pm$ 0.29 & 0.204\cr
\noalign{\medskip}
(2,2) & 714 $\pm$ 27 & 362 $\pm$ 102 & -- & 1.52 $\pm$ 0.30  & 0.194\cr
\noalign{\medskip}
\multispan5{Breit-Wigner fit with constant background\hfil}\cr
\noalign{\medskip}
(1,1) & 746$\pm$16 & 145 $\pm$ 69 & 0.18 $\pm$ 0.10 & 0.59 $\pm$  
0.18 & 0.712\cr
\noalign{\medskip}
(1,2) & 706 $\pm$ 39 & 262 $\pm$ 24 & 0.13 $\pm$ 0.39 & 1.05 $\pm$  
0.44 & 0.114\cr
\noalign{\medskip}
(2,1) & 745 $\pm$ 30 & 165 $\pm$ 112 & 0.23 $\pm$ 0.21 & 0.97 $\pm$  
0.42 & 0.094\cr
\noalign{\medskip}
(2,2) & 724 $\pm$ 25 & 211 $\pm$ 117 & 0.25 $\pm$ 0.20 & 1.41 $\pm$  
0.42 & 0.124\cr}
\bigskip
\bigskip
\noindent
{\bf Table 9.} Results of the fits to the four solutions of the  
$S$-wave intensity measured in $\pi^+ n \to \pi^+ \pi^- p$ at 5.98  
GeV/c using Monte Carlo method for amplitude analysis. The fits are  
made with Breit-Wigner parametrization (6.3) and (6.4) with the Pi\v  
s\' ut-Roos shape factor.
\vfill\eject

\halign to \hsize{\hfil #\hfil & \hfil #\hfil & \hfil #\hfil &  
\hfil # \hfil & \hfil # \hfil & \hfil #\hfil\cr
$I_S$ & $m_\sigma$ & $\Gamma_\sigma$ & $B$ & $N_S$ & $\chi^2$/dof\cr
Solution & (MeV) & (MeV)\cr
\noalign{\medskip\hrule\medskip}
\multispan3{Single Breit-Wigner fit\hfil}\cr
\noalign{\medskip}
(1,1) & 778 $\pm$ 10 & 158 $\pm$ 21 & -- & 1.19 $\pm$ 0.12 & 2.158\cr
\noalign{\medskip}
(1,2) & 749 $\pm$ 31 & 353 $\pm$ 88 & -- & 1.47 $\pm$ 0.25 & 0.430\cr
\noalign{\medskip}
(2,1) & 752 $\pm$ 20 & 237 $\pm$ 52 & -- & 1.50 $\pm$ 0.27 & 0.844\cr
\noalign{\medskip}
(2,2) & 749 $\pm$ 19 & 309 $\pm$ 63 & -- & 1.92 $\pm$ 0.24  & 0.632\cr
\noalign{\medskip}
\multispan5{Breit-Wigner fit with constant background\hfil}\cr
\noalign{\medskip}
(1,1) & 782$\pm$9 & 117 $\pm$ 26 & 0.08 $\pm$ 0.03 & 1.13 $\pm$  
0.16 & 1.024\cr
\noalign{\medskip}
(1,2) & 770 $\pm$ 24 & 202 $\pm$ 74 & 0.09 $\pm$ 0.04 & 1.52 $\pm$  
0.32 & 0.080\cr
\noalign{\medskip}
(2,1) & 763 $\pm$ 18 & 153 $\pm$ 55 & 0.12 $\pm$ 0.05 & 1.46 $\pm$  
0.39 & 0.236\cr
\noalign{\medskip}
(2,2) & 756 $\pm$ 15 & 200 $\pm$ 59 & 0.11 $\pm$ 0.05 & 1.93 $\pm$  
0.33 & 0.212\cr}
\bigskip
\bigskip
\noindent
{\bf Table 10.} Results of the fits to the four solutions of the  
$S$-wave intensity measured in $\pi^+ n \to \pi^+ \pi^- p$ at 11.85  
GeV/c using Monte Carlo method for amplitude analysis. The fits are  
made with Breit-Wigner parametrization (6.3) and (6.4) with the Pi\v  
s\' ut-Roos shape factor.
\vfill\eject

\noindent
{\bf Figure Captions.}
\bigskip
\bigskip
\noindent
{\bf Fig.~1.} Mass dependence of unnormalized amplitudes  
$|\overline S |^2\Sigma$ and $|S|^2\Sigma$ measured in $\pi^-  
p_\uparrow \to \pi^- \pi^+ n$ at 17.2 GeV/c at $-t = 0.005 - 0.20$  
(GeV/c)$^2$ using the Monte Carlo method for amplitude analysis  
(Ref.~23). Both solutions for the amplitude $|\overline S|^2\Sigma$  
resonate at 750 MeV while the amplitude $|S|^2\Sigma$ is  
nonresonating in both solutions.
\bigskip
\noindent
{\bf Fig.~2.} Mass dependence of unnormalized amplitudes  
$|\overline S|^2\Sigma$ and $|S|^2\Sigma$ measured in $\pi^-  
p_\uparrow \to \pi^- \pi^+ n$ at 17.2 GeV/c at $-t = 0.005 - 0.20$  
(GeV/c)$^2$ using the $\chi^2$ minimization method for amplitude  
analysis. Based on Fig.~10 of Ref.~13 and Fig.~VI-21 of Ref.~12.  
Both solutions for the amplitude $|\overline S|^2\Sigma$ resonate at  
750 MeV while the amplitude $|S|^2\Sigma$ is nonresonating in both  
solutions. The analysis used the same data as in Fig.~1 (20 MeV mass  
bins).
\bigskip
\noindent
{\bf Fig.~3.} The fits to amplitude $|\overline S|^2\Sigma$ using  
the single Breit-Wigner parametrization (4.1) with Pi\v s\' ut-Roos  
shape factor (4.2a). The fitted parameters are given in Table 1.
\bigskip
\noindent
{\bf Fig.~4.} The fits to amplitude $|\overline S|^2\Sigma$ using  
the single Breit-Wigner parametrization (4.1) with phenomenological  
shape factor $F = 1$. The fitted parameters are given in Table 1.
\bigskip
\noindent
{\bf Fig.~5.} The fits to amplitude $|\overline S|^2\Sigma$ using  
the Breit-Wigner parametrization (4.6) with constant incoherent  
background and with phenomenological shape factors $F=1$. The fitted  
parameters are given in Table 2.
\bigskip
\noindent
{\bf Fig.~6.} The fits to amplitude $|\overline S|^2\Sigma$ using  
the Breit-Wigner parametrization (4.15) with constant coherent  
background and with phenomenological shape factor $F=1$. The fitted  
parameters are given in Table 3.
\bigskip
\noindent
{\bf Fig.~7.} The fits to amplitude $|\overline S|^2\Sigma$ using  
the Breit-Wigner parametrization (4.19) with $t$-averaged constant  
coherent background and with phenomenological shape factor $F=1$.  
The fitted parameters are given in Table 4.
\bigskip
\noindent
{\bf Fig.~8.} Mass dependence of unnormalized amplitudes  
$|\overline S|^2\Sigma$ and $|S|^2\Sigma$ measured in $\pi^- p \to  
\pi^- \pi^+ n$ at 17.2 GeV/c at $-t = 0.005 - 0.20$ (GeV/c)$^2$ in  
40 MeV mass bins from 600 to 1520 MeV. Based on Figs.~2 and 6 from  
Ref.~14. The amplitude $|\overline S|^2\Sigma$ resonates at 750 in  
Solution 1 and at 800 MeV in Solution 2 while the amplitude for  
$|S|^2\Sigma$ is nonresonating in both solutions in this mass range.
\bigskip
\noindent
{\bf Fig.~9.} The fits to amplitude $|\overline S|^2\Sigma$ using  
the Breit-Wigner parametrization (5.4) below 1120 MeV and a single  
Breit-Wigner formula with incoherent background above 1120 MeV. The  
phenomenological shape factor $F=1$. The fitted parameters are given  
in Tables 5 and 6.
\bigskip
\noindent
{\bf Fig.~10.} Four solutions for the $S$-wave intensity $I_S$  
measured in the reaction $\pi^- p_\uparrow \to \pi^- \pi^+ n$ at  
17.2 GeV/c and $-t = 0.005 -0.2$ (GeV/c)$^2$ using Monte Carlo  
method for amplitude analysis (Ref.~23). The solid curves are fits  
to single Breit-Wigner parametrization (6.3). The dashed curves are  
fits to Breit-Wigner parametrization (6.4) with incoherent  
background. The fitted parameters are given in Table 7.
\bigskip
\noindent
{\bf Fig.~11.} Two of the four solutions for the $S$-wave intensity  
$I_S$ measured in the $\pi^- p_\uparrow \to \pi^-\pi^+ n$ at 17.2  
GeV/c and $-t=0.005 - 0.2$ (FeV/c)$^2$ using $\chi^2$ minimization  
method for amplitude analysis. The data are based on Fig.~14a of  
Ref.~13 and Fig.~12 of Ref.~11. The solid and dashed curves are  
Breit-Wigner fits as in Fig.~10. The fitted parameters are given in  
Table 8.
\bigskip
\noindent
{\bf Fig.~12.} Four solutions for the $S$-wave intensity $I_S$  
measured in $\pi^+ n_\uparrow \to \pi^+ \pi^- p$ at 5.98 GeV/c and  
$-t = 0.2 - 0.4$ (GeV/c)$^2$ using Monte Carlo method for amplitude  
analysis (Ref.~23). The solid and dashed curves are Breit-Wigner  
fits as in Fig.~10. The fitted parameters are given in Table 9.
\bigskip
\noindent
{\bf Fig.~13.} Four solutions for the $S$-wave intensity $I_S$  
measured in $\pi^+ n_\uparrow \to \pi^+ \pi^- p$ at 11.85 GeV/c and  
$-t = 0.2 - 0.4$ (GeV/c)$^2$ using Monte Carlo method for amplitude  
analysis (Ref.~23). The solid and dashed curves are Breit-Wigner  
fits as in Fig.~10. The fitted parameters are given in Table 10.
\bigskip
\noindent
{\bf Fig.~14.} The $t$-evolution of mass dependence of moduli  
squared of $t$-channel normalized transversity amplitudes $|L|^2$,  
$|\overline L|^2$, $|U|^2$ and $|\overline U|^2$ in $\pi^+  
n_\uparrow \to \pi^+\pi^- p$ at 5.98 GeV/c together with results for  
$\pi^- p_\uparrow \to \pi^- \pi^+ n$ at 17.2 GeV/c and $t=0.068$  
(GeV/c)$^2$.
\bigskip
\noindent
{\bf Fig.~15.} The ratio of amplitudes with recoil nucleon  
transversity ``down'' and ``up'' with dimeson helicity $\lambda =  
0$. The deviation from unity shows the strength of $A_1$-exchange  
amplitudes. Based on Fig.~6 of Ref.~14. In our notation, $g_S = S$,  
$h_S = \overline S$, $g_P = L$, $h_P = \overline L$.
\bigskip
\noindent
{\bf Fig.~16.} Test of predictions $\rho^y_{ss} + \rho^y_{00} +  
2\rho^y_{11} = -2 (\rho^y_{00} - \rho^y_{11}) = + 2\rho^y_{1-1}$ due  
to vanishing $A_1$-exchange in $\pi^- p_\uparrow \to \pi^- \pi^+ n$  
at 17.2 GeV/c and $-t = 0.005 - 0.2$ (GeV/c)$^2$.
\bigskip
\noindent
{\bf Fig.~17.} Test of predictions ${\rm Re}\rho^y_{10} = {\rm  
Re}\rho^y_{1s} = {\rm Re} \rho^y_{0s} = 0$ due to vanishing  
$A_1$-exchange in $\pi^- p_\uparrow \to \pi^- \pi^+ n$ at 17.2 GeV/c  
and $-t = 0.005 - 0.2$ (GeV/c)$^2$.
\bigskip
\noindent
{\bf Fig.~18.} $S$-wave intensity normalized to 1 at maximum value.  
The data correspond to solutions $I_S(1,1)$ and $I_S(2,2)$ at 17.2  
(GeV/c) from Ref.~23. The smooth curves are predictions of phase  
shift analysis for $\pi^+ \pi^- \to \pi^+ \pi^-$ from Ref.~27. The  
dashed curve is the accepted solution Down, the dot-dashed curve is  
the rejected solution Up.
\bye